\documentclass[fleqn,usenatbib]{mnras}

\usepackage{newtxtext,newtxmath}
\usepackage[T1]{fontenc}

\DeclareRobustCommand{\VAN}[3]{#2}
\let\VANthebibliography\thebibliography
\def\thebibliography{\DeclareRobustCommand{\VAN}[3]{##3}\VANthebibliography}

\usepackage{graphicx}	% Including figure files
\usepackage{amsmath}	% Advanced maths commands
\title[\textbf{Nulling and drifting in PSR J1727$-$2739}]{\textbf{Nulling and subpulse
  drifting in PSR J1727$-$2739}}

\author[R. Rejep et al.]{
  Rukiye Rejep,$^{1,2}$
  N. Wang,$^{1,3,4}$
  W. M. Yan$^{1,3,4}$\thanks{E-mail: yanwm@xao.ac.cn (WMY)}
  and
  Z. G. Wen$^{1,3,4}$
\\
$^{1}$Xinjiang Astronomical Observatory, CAS, 150 Science 1-Street, Urumqi, Xinjiang 830011, China\\
$^{2}$University of Chinese Academy of Sciences, Beijing 100049, China\\
$^{3}$Key Laboratory of Radio Astronomy, Chinese Academy of Sciences, Nanjing 210008, China\\
$^{4}$Xinjiang Key Laboratory of Radio Astrophysics, 150
Science 1-Street, Urumqi, Xinjiang
830011, China
}

\date{Accepted 2021 October 17. Received October 15; in original form 2021 July 07}

\pubyear{2021}

\begin{document}
\label{firstpage}
\pagerange{\pageref{firstpage}--\pageref{lastpage}}
\maketitle

\begin{abstract}
  
  In this paper, we investigate the emission properties of PSR J1727$-$2739, whose mean pulse
  profile has two main components, by analysing five
single-pulse observations made using the Parkes 64-m radio telescope with a central
frequency of 1369 MHz between 2014 April and October. The total
observation time is about 6.1 hours which contains 16718 pulses after removal of radio
frequency interference (RFI). Previous studies reveal that PSR J1727$-$2739 exhibits both
nulling and subpulse drifting. We estimate the nulling fraction to be
66\%$\pm$1.4\%, which is consistent with previously published results.
In addition to the previously known
subpulse drifting in the leading component, we also explore the
drifting properties for the trailing component.
  We observe two distinct drift modes whose vertical drift band
  separations ($P_{3}$) are consistent with earlier studies.
  We find that both profile components share the same drift periodicity
  $P_{3}$ in a certain drift mode, but the measured horizontal separations
  ($P_{2}$) are quite different for them. That is, PSR J1727$-$2739 is
  a pulsar showing both changes of drift periodicity $P_{3}$
  between different drift modes and drift rate variations between
  components in a given drift mode.
Pulsars exhibiting nulling along with drift mode changing,
such as PSR J1727$-$2739, give an unique opportunity to
investigate the physical mechanism of these phenomena.

\end{abstract}

\begin{keywords}
  stars: neutron -- pulsars: general -- pulsars: individual (PSR J1727$-$2739)
\end{keywords}

%%%%%%%%%%%%%%%%%%%%%%%%%%%%%%%%%%%%%%%%%%%%%%%%%%

%%%%%%%%%%%%%%%%% BODY OF PAPER %%%%%%%%%%%%%%%%%%

\section{Introduction}

The radio emission of pulsars usually remains stable in intensity
and polarization. However, some pulsars show remarkable emission variations,
such as nulling, mode-changing, and subpulse drifting. Pulse nulling is
the abrupt cessation of pulsed radio emission for a number of spin periods
\citep{rit76,ran86,wmj07,gjk12}. The degree of nulling in a pulsar is known
as the nulling fraction (NF), which is the fraction of pulses with no detectable
emission. The observed NF for nulling pulsars ranges from less than 1\% to more
than 95\% \citep{wmj07,gjw14,gyy+17,whh+20}. This phenomenon was first discovered by
\citet{bac70}. To date, pulse nulling has been reported in more than 200
pulsars \citep{whh+20}.
For some pulsars, their mean pulse profiles switch between
two or more quasi-stable emission states. This phenomenon is referred to as mode
changing. It was first observed in PSR J1239+2453 \citep{bac70a}. The typical
time-scale of nulling and mode changing is several seconds to hours. In some cases, 
transitions between emission states are periodic. Periodic nulling has been observed 
in many pulsars \citep{hr07,hr09,rw08,rwb13,gjw14,gyy+17,bmm17,bm18,bpm19,bm19}.
Recently, periodic mode changing also has been reported in some pulsars
\citep{mr17,bm19,ymw+19,ymw+20}.

Another well-known emission variation
is subpulse drifting. Subpulse drifting is the regular phase drifting of subpulses.
Since it was discovered by \citet{dc68}, subpulse drifting has been observed in about
120 pulsars \citep{wes06,wse07,bmm+16,bmm+19}.
The drift patterns are separated horizontally across the pulse phase by $P_{2}$
(in units of degree) and vertically across time by $P_{3}$ (in units of pulse period $P_{1}$).
The drift rate $\bigtriangleup\phi$ which represents the drift speed is
then given by: $\bigtriangleup\phi = P_{2}/P_{3}$ (in units of \degr/$P_{1}$)
\citep{smk05,wwy+16}.

Many studies reveal that nulling, mode changing and subpulse drifting may be closely
linked to each other. Some pulsars show evidence of interaction between nulling and
subpulse drifting. For example, the subpulse of PSR B0809+74 was found to drift
more slowly after the null states \citep{vsrr03}.
\citet{jv04} reported that, for PSR B0818$-$13, the subpulse drift
appears to speed up during the nulls. \citet{gyy+17} studied the nulling--drifting
interaction in PSR J1840$-$0840. They found that this pulsar tends to start nulling
after the end of a drift band. Then, when the pulsar switches back on, it often starts
at the beginning of a new drift band in both pulse profile components. Studies of the
interaction between nulling and subpulse drifting can provide insights into the
properties of pulsar magnetosphere \citep{rwr05,fr10}. \citet{wmj07} show clear
evidence of a close relationship between nulling and mode changing. It is suggested 
that nulling and mode changing are different manifestations of the same physical process.

PSR J1727$-$2739 was discovered by \citet{hfs+04} in the Parkes Multibeam Pulsar
Survey. This pulsar has a characteristic age of $1.86\times10^{7}$ yr and a spin
period of 1.29~s. It is known to show both nulling and subpulse drifting. 
\citet{wwy+16} studied nulling and subpulse drifting properties
for PSR J1727$-$2739 using a 2-hr observation. They found that this pulsar shows nulls
with lengths lasting from 6 to 281 pulses and separated by burst phases ranging from
2 to 133 pulses, and they measured a NF of 68\%. Two distinct subpulse drift modes
were identified, with vertical spacing between the drift bands $P_{3}$ of 9.7 $\pm$ 1.6
and 5.2 $\pm$ 0.9 pulse periods, respectively.
In this paper, we carry out a detailed investigation of the single-pulse behaviour of
PSR J1727$-$2739 with observations lasting for 6.1 hr. The observations and
data processing are given in Section~\ref{sec:obs}.
Details of the results are presented in Section~\ref{sec:results}.
The final conclusions and implications of our results
are discussed in Section~\ref{sec:discussion}.

\section{Observations}\label{sec:obs}

The observational data analysed here were downloaded from the Parkes Pulsar Data
Archive that is freely available online\footnote{\url{https://data.csiro.au}}
\citep{hmm+11}. The observations  were carried out using the Parkes 
64-m radio telescope from April 1 to October 27 in 2014 at 5 epochs (see
Table~\ref{tab:nf} for exact observing dates) with the center
beam of the multibeam receiver \citep{swb+96} and the Parkes digital filterbank
systems PDFB3 and PDFB4 \citep{mhb+13}. The total bandwidth was 256 MHz centred
at 1369 MHz with 512 channels across the band.
For each observation, the data were sampled every 256 $\mu$s.
The total integration time is 6.1 hr for all observations.

We used the {\tt\string DSPSR} package \citep{vb11} to de-disperse and fold the
data with the ephemeris from the ATNF Pulsar Catalogue
V1.64\footnote{\url{http://www.atnf.csiro.au/research/pulsar/psrcat/}}
\citep{mhth05}. The single-pulse integrations were produced, which were recorded
using the {\tt\string PSRFITS} data format
\citep{hvm04} with 1024 phase bins per rotation period. Band edges (5 per cent
on each side) and strong narrow-band and impulsive RFI were removed from the
archive files using {\tt\string PAZ} and {\tt\string PAZI} programs of the
{\tt\string PSRCHIVE} packages \citep{hvm04}. After the RFI-excision procedure,
a total of 16718 pulses were obtained from all observations.
Then the RFI excised data were processed using the {\tt\string PSRCHIVE} packages.
The analysis of fluctuation spectra was performed with the {\tt\string PSRSALSA}
package \citep{wel16} which is available 
online\footnote{\url{https://github.com/weltevrede/psrsalsa}}. We followed \citet{ymv+11}
to carry out polarization calibration using the {\tt\string PSRCHIVE}
program {\tt\string PAC}.

\section{Results}\label{sec:results}

In this section, we present details of the nulling and subpulse drifting of 
PSR J1727$-$2739.

\subsection{Nulling}\label{sec:null}

\begin{figure}
\centering
\includegraphics[angle=0,width=\columnwidth]{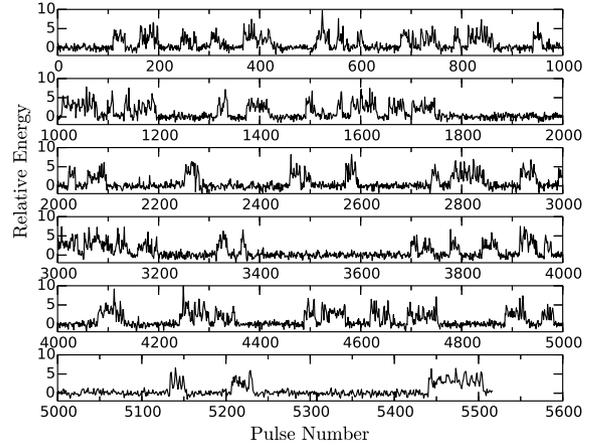}
\caption{Pulse energy variations for the 2014-04-01
    observation (see Table~\ref{tab:nf}
  for details). Note that pulses
  contaminated by RFI have been clipped. The total number of pulses displayed
  in this plot is 5444.
        \label{fig:energy_var} }
\end{figure}

\begin{figure*}
  \begin{tabular}{cccccc}
\includegraphics[width=2.2in,angle=0]{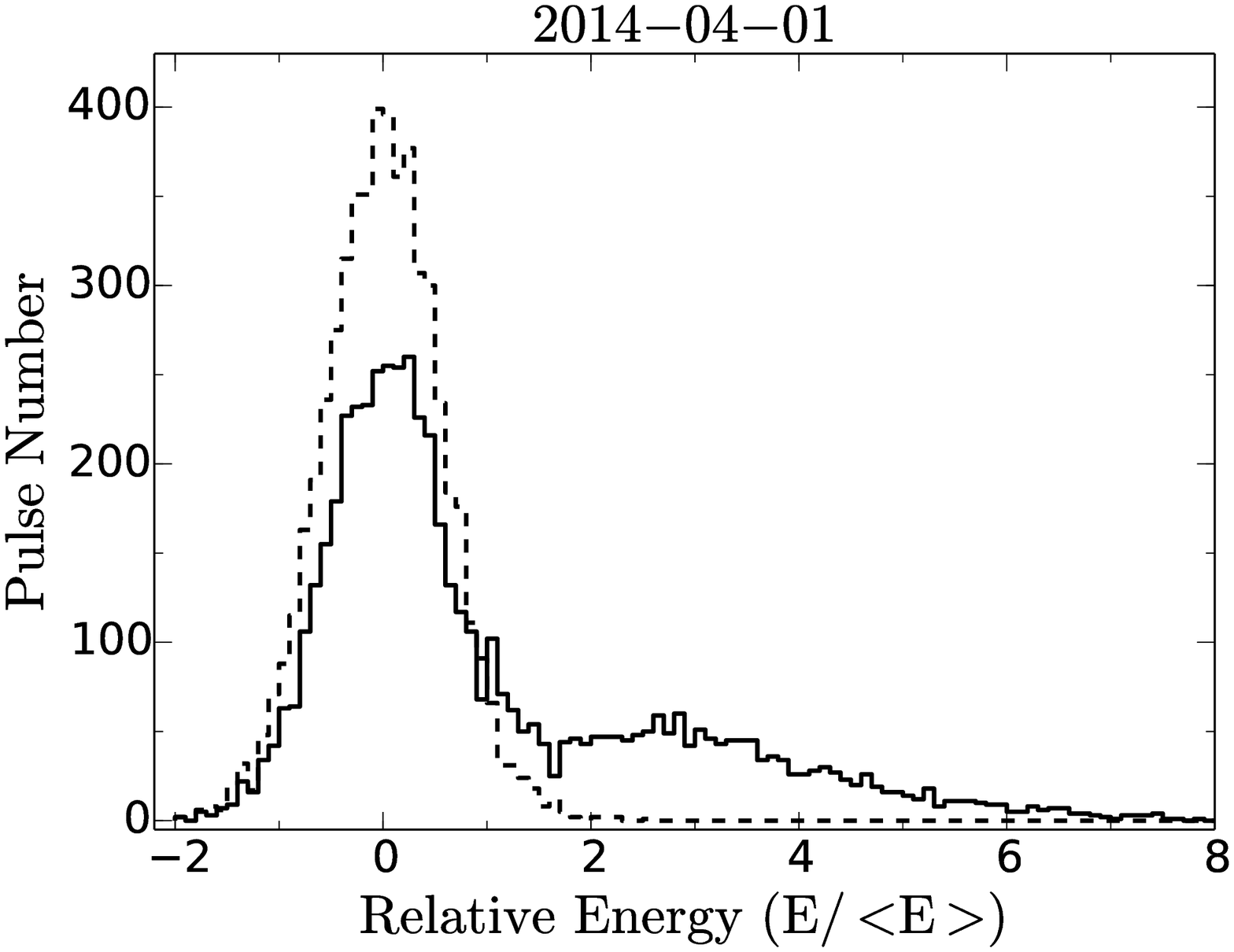}    
&\includegraphics[width=2.2in,angle=0]{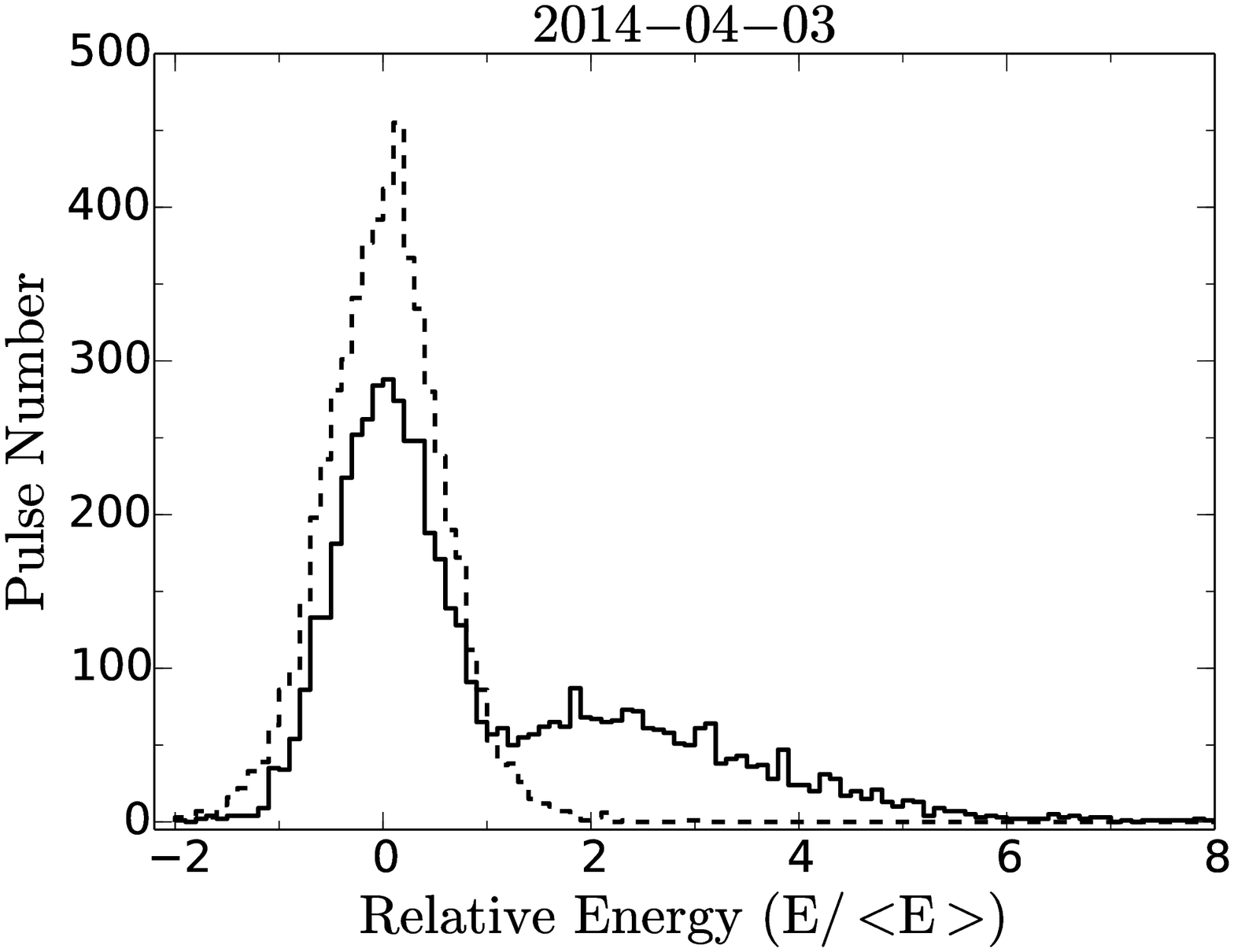}
&\includegraphics[width=2.2in,angle=0]{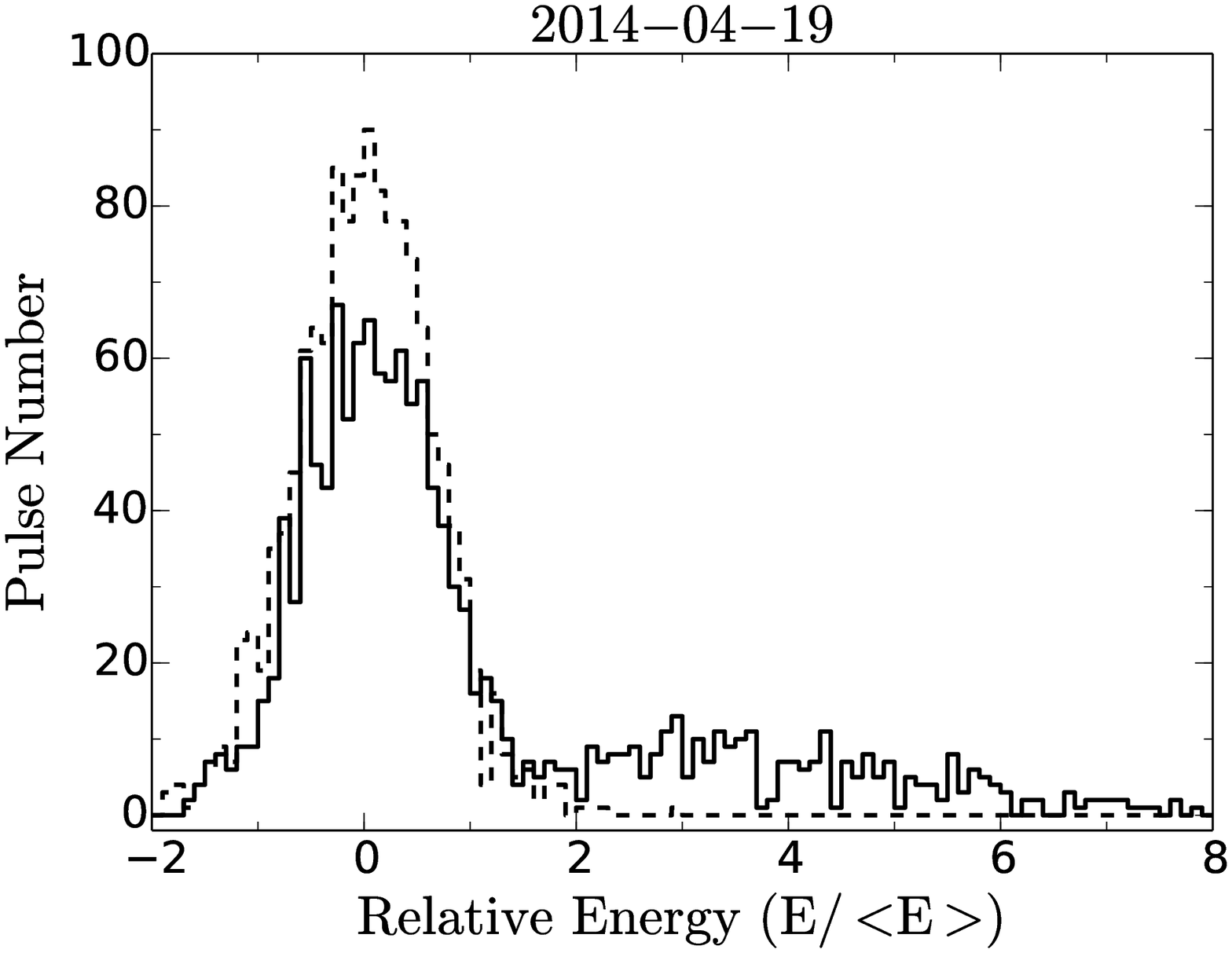}\\
\includegraphics[width=2.2in,angle=0]{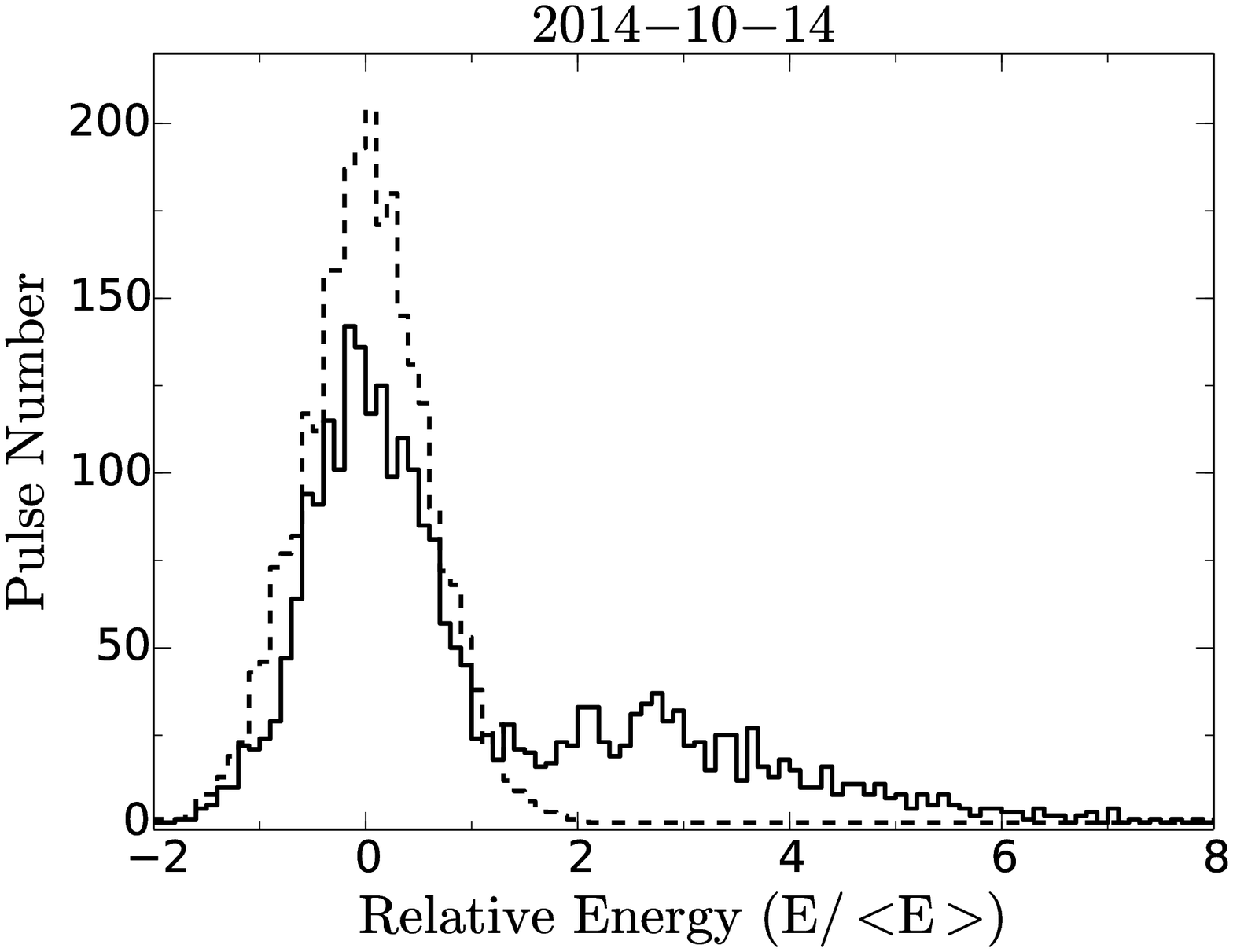}
&\includegraphics[width=2.2in,angle=0]{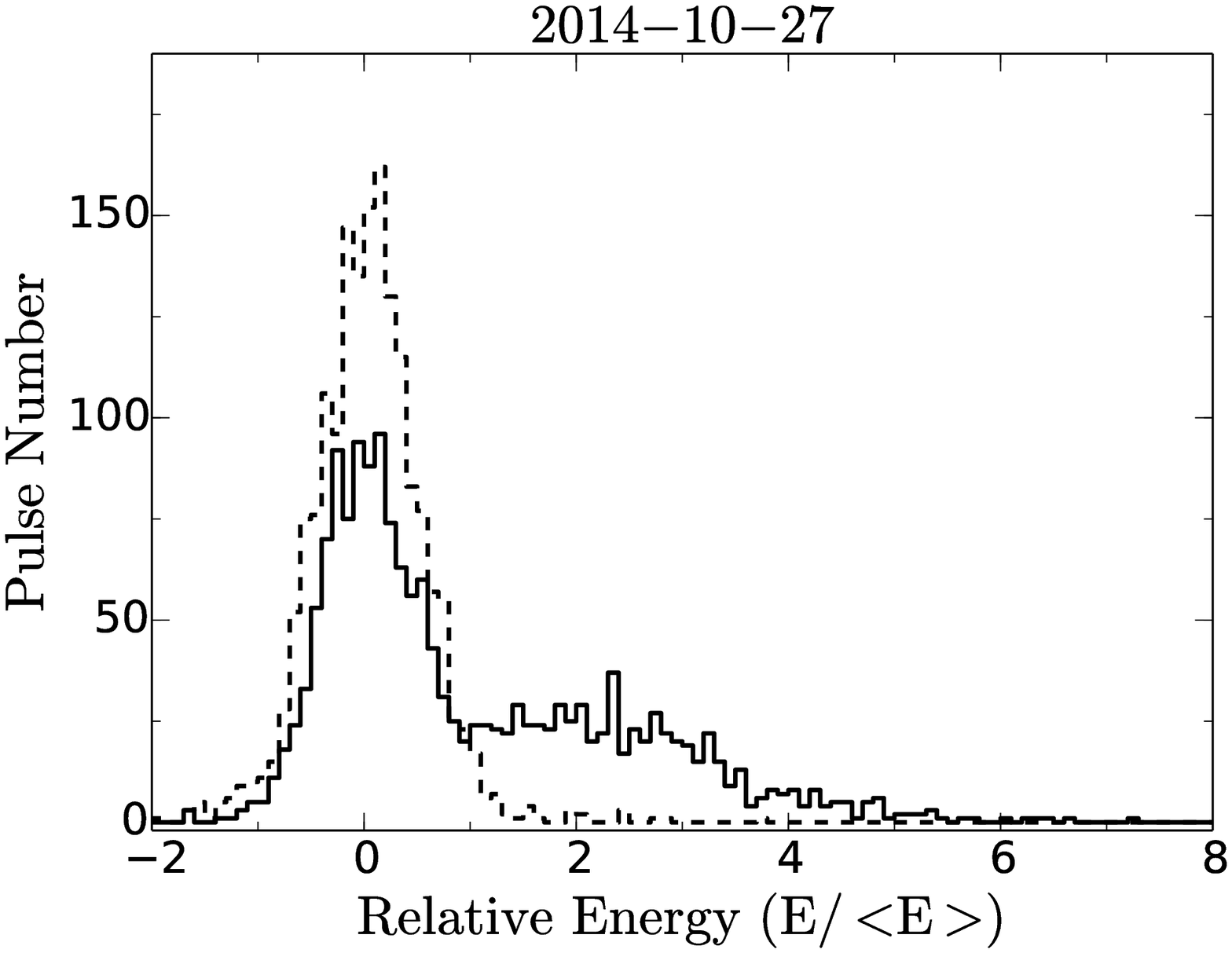}\\
\end{tabular}
\caption{Pulse energy distributions for the on-pulse (solid histogram) and off-pulse
  (dashed histogram) regions for all observations (see Table~\ref{tab:nf}
  for details).
        \label{fig:energy_dist} }
 \end{figure*}

\begin{table}
\centering
\caption{Nulling fractions for 5 observations of PSR J1727$-$2739. Note that the symbols
  $\rm T_{obs}$, $\rm N_{tot}$ and $\rm N_{null}$ represent the duration, the total number
of pulses and the number of null pulses of the observation,
respectively.  RMS$_{\rm off}$ in the last column is the mean rms of
the single-pulse baseline noise.}
\label{tab:nf}
\begin{tabular}{ccccccc}
\hline
Date  &$\rm T_{obs}$  &$\rm N_{tot}$  &$\rm N_{null}$  &NF  &RMS$_{\rm
                                                              off}$\\
(yyyy-mm-dd)  &(min)   &   &  &(\%)  &\\
\hline
2014-04-01    & 119  & 5444  &3674    &$66\pm1.1$  &$1.55\pm0.06$\\
2014-04-03    & 120  & 5522  &3546    &$64\pm1.1$  &$1.31\pm0.04$\\
2014-04-19    & 30   & 1356  &1015    &$73\pm2.3$  &$1.77\pm0.06$\\
2014-10-14    & 58   & 2683  &1819    &$69\pm1.6$  &$1.58\pm0.06$\\
2014-10-27    & 39   & 1713  &1022    &$62\pm1.9$  &$1.21\pm0.05$\\
\hline
\end{tabular}
\end{table}

\begin{figure*}
  \begin{tabular}{cccccc}
\includegraphics[width=2.2in,angle=0]{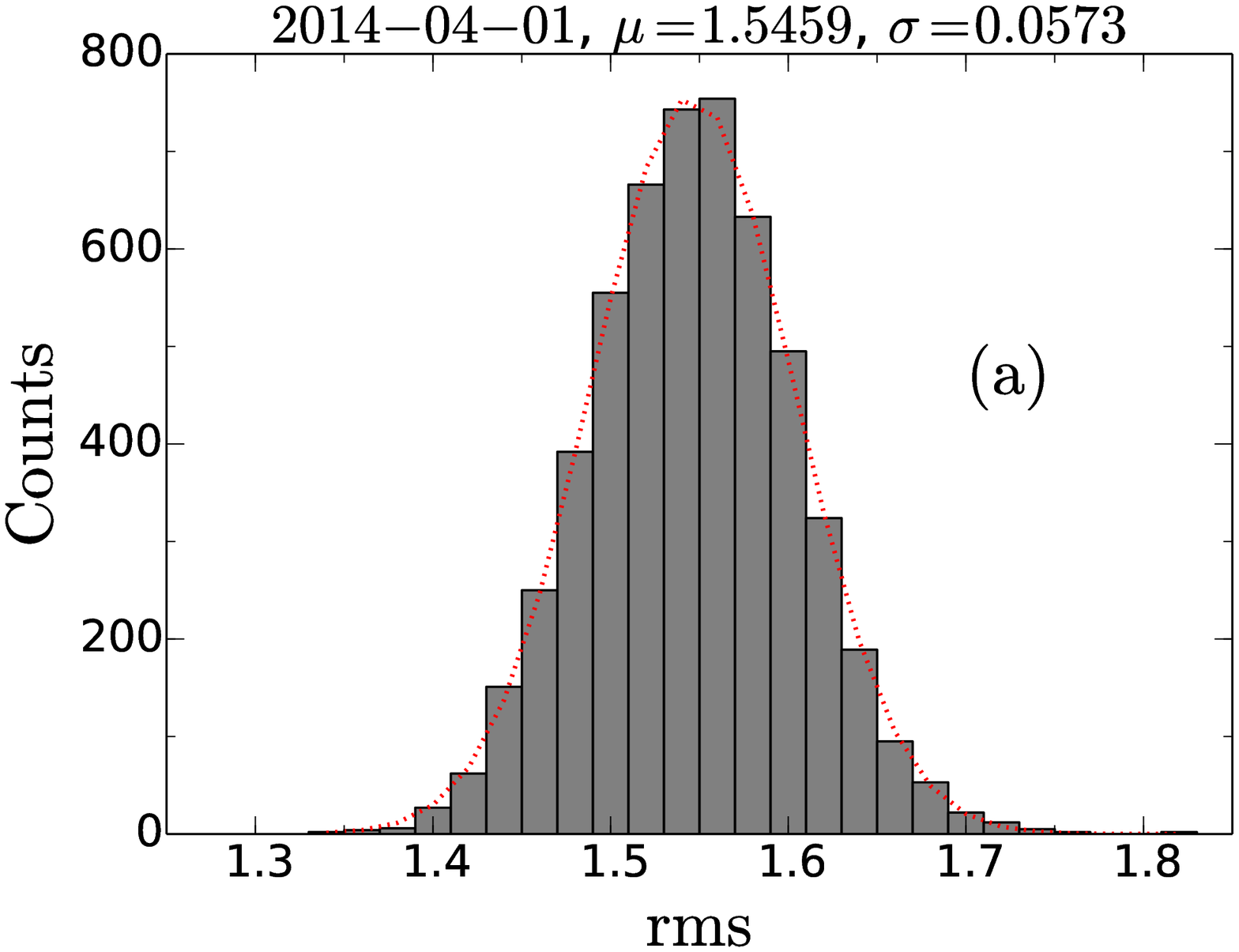}    
&\includegraphics[width=2.2in,angle=0]{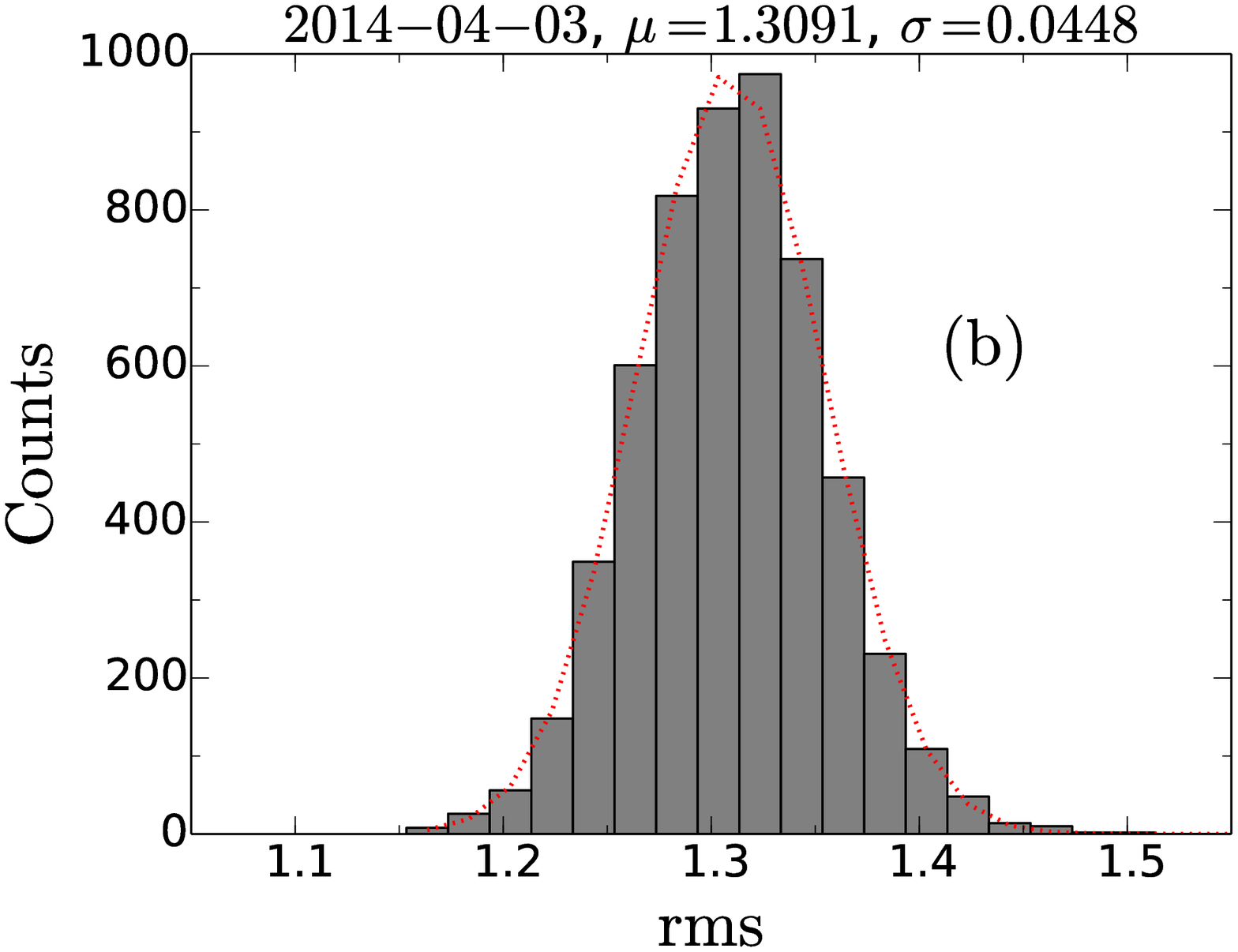}
&\includegraphics[width=2.2in,angle=0]{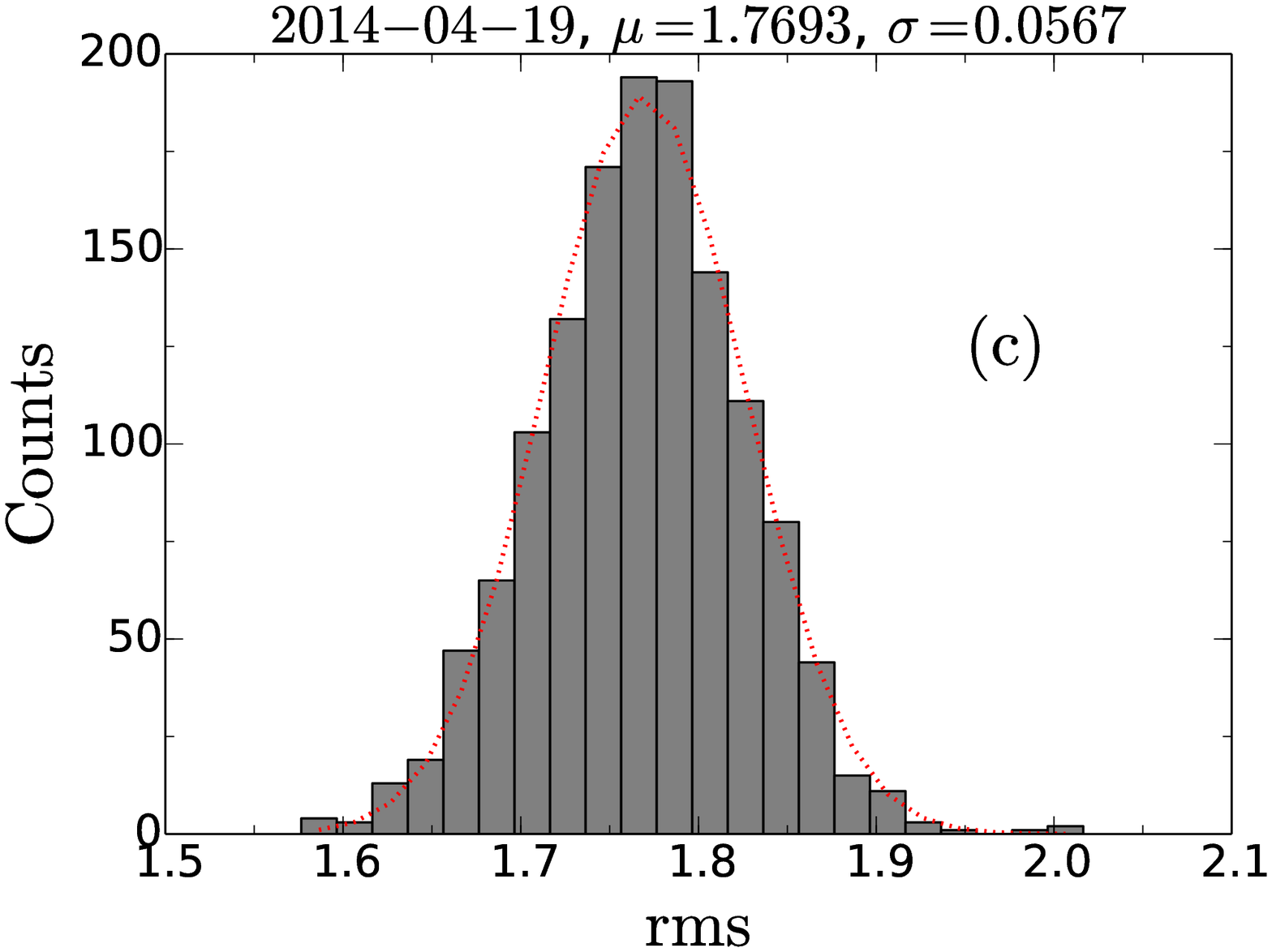}\\
\includegraphics[width=2.2in,angle=0]{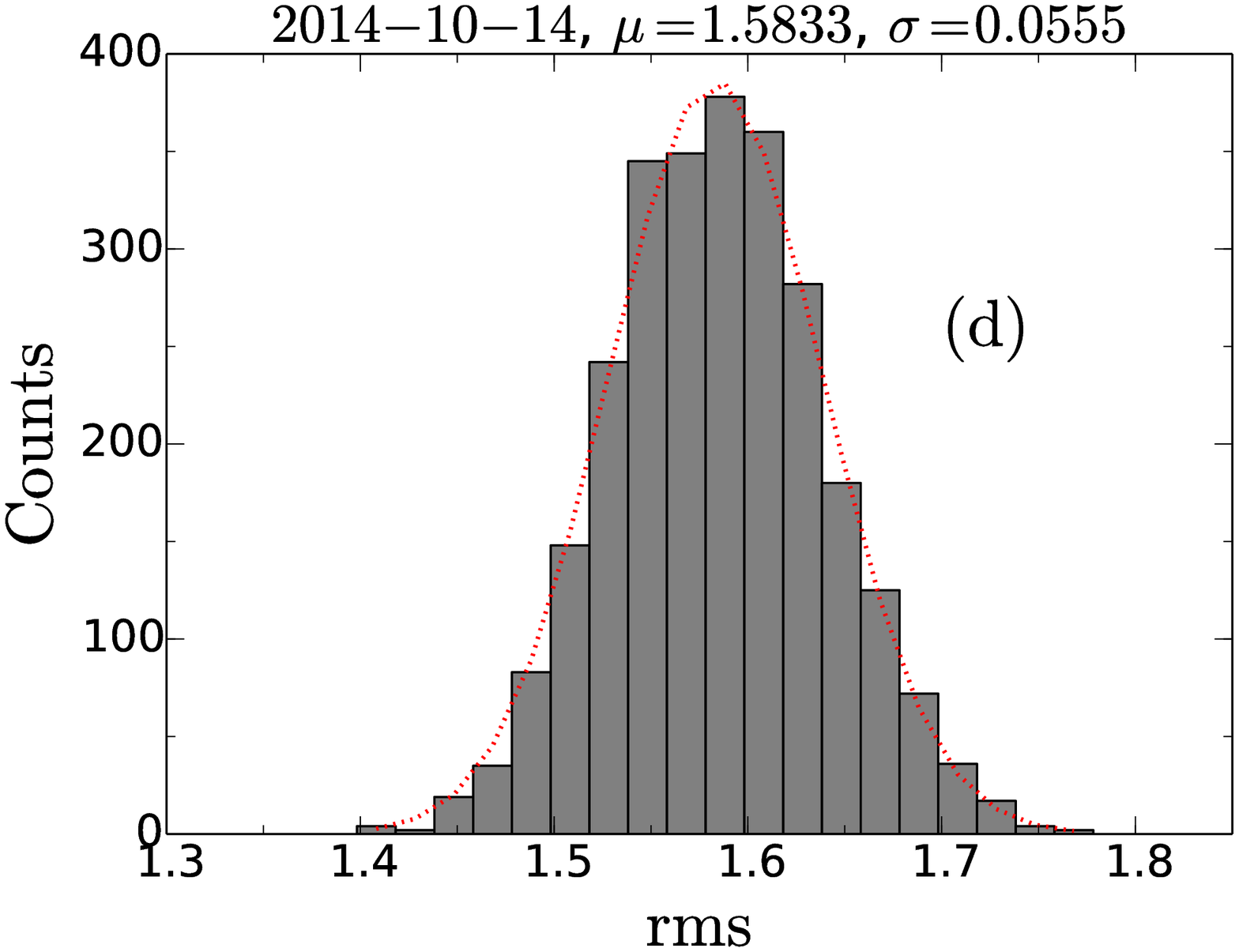}
&\includegraphics[width=2.2in,angle=0]{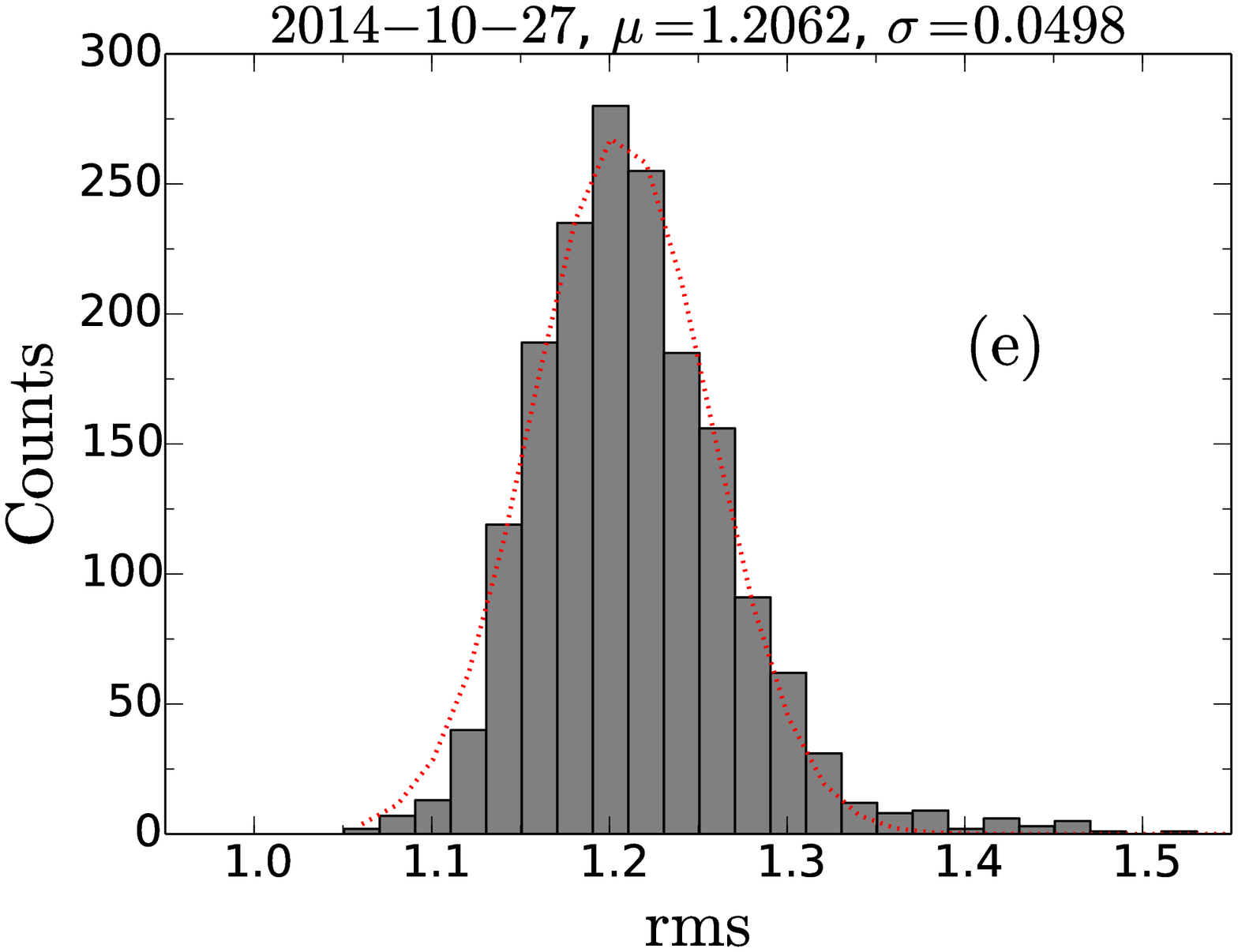}
&\includegraphics[width=2.2in,angle=0]{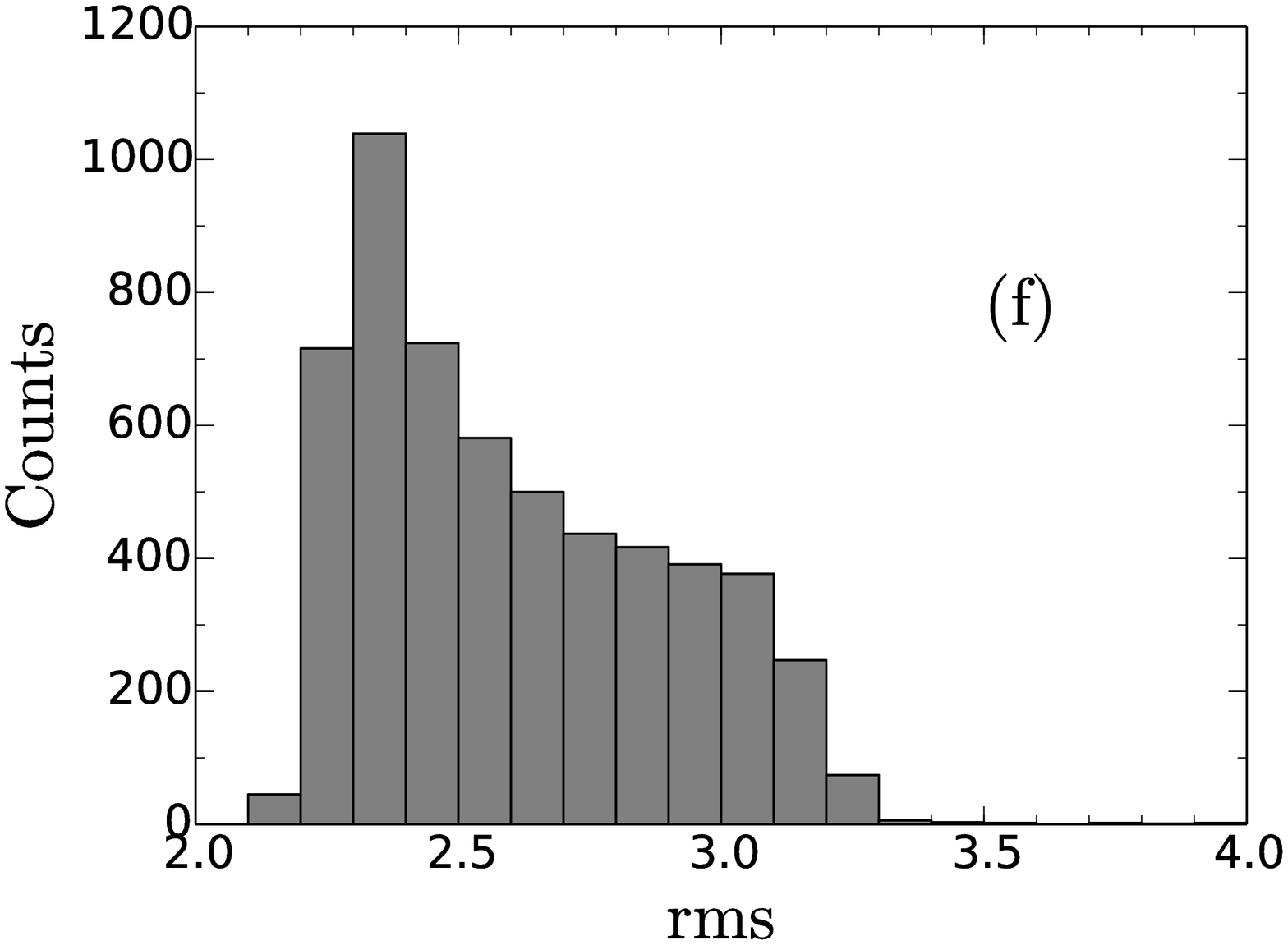}\\
\end{tabular}
\caption{Distributions of the rms of the baseline noise for observations
    of this study (panels (a) to (e)) and \citet{wwy+16} (panel (f)). The red
    dotted lines are Gaussian distributions that have been fitted to the rms
    histograms.
}\label{fig:rms_dist}
 \end{figure*}

\begin{figure*}
\begin{tabular}{cc}
\includegraphics[width=\columnwidth,angle=0.0]{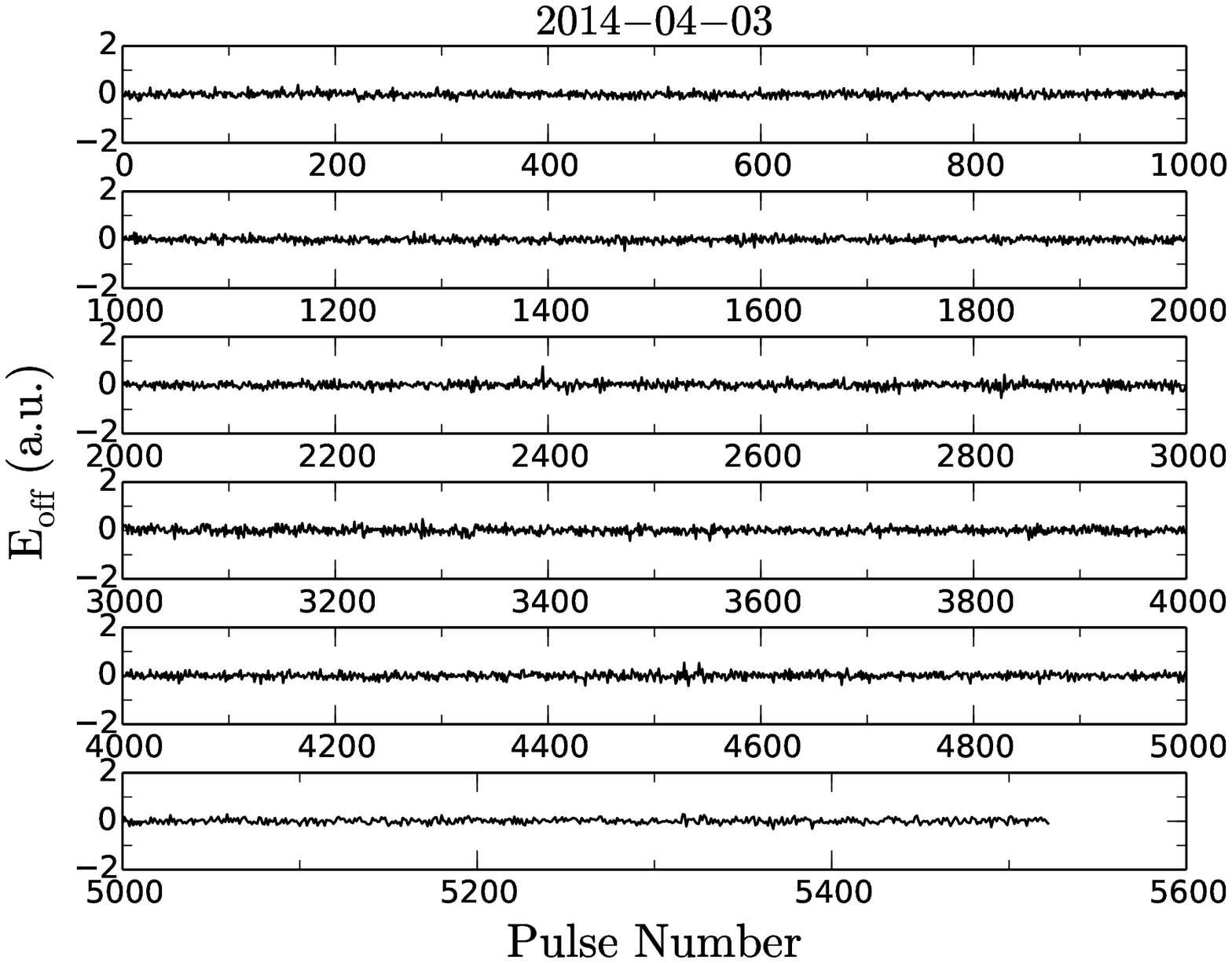}
&\includegraphics[width=\columnwidth,angle=0.0]{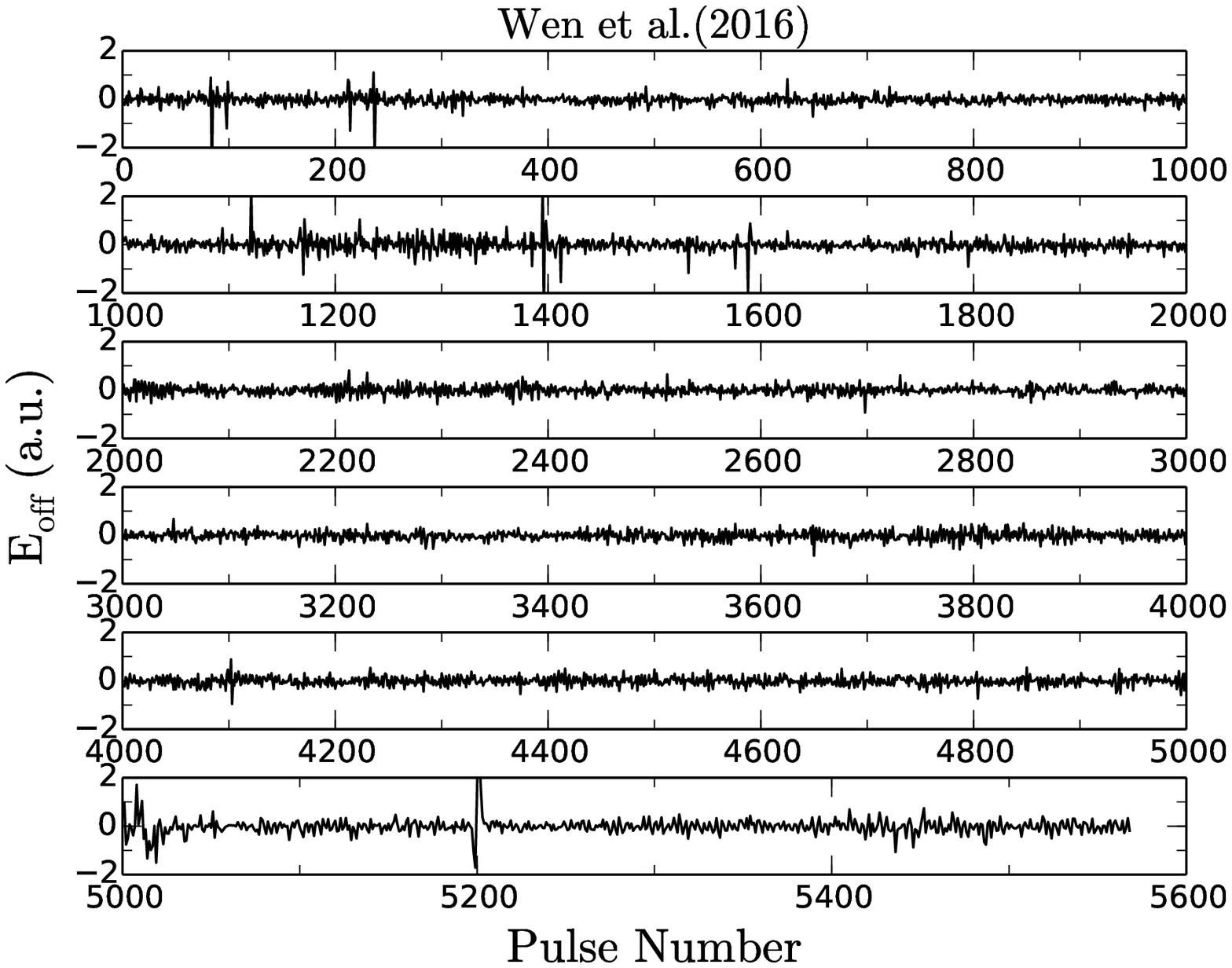}\\
\end{tabular}
\caption{Off-pulse energy variations with time for the 2014-04-03
  observation (left) and the observation of \citet{wwy+16} (right).
}\label{fig:eoff_var}
\end{figure*}

\begin{figure}
\begin{tabular}{ccc}
\includegraphics[width=\columnwidth,angle=0]{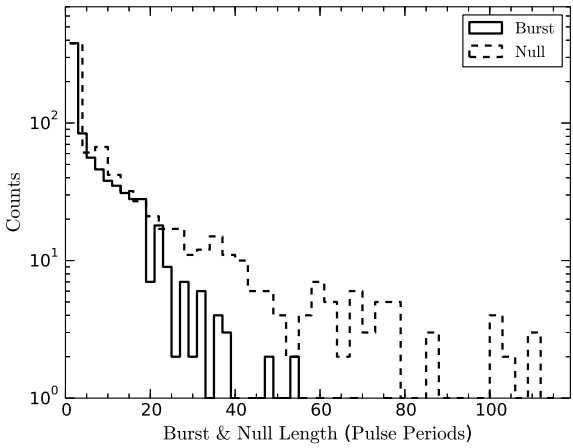}
\end{tabular}
\caption{Distributions of the durations of the burst (solid line) and null
  (dashed line) states. 
}\label{fig:null_len}
\end{figure}

\begin{figure}
\begin{tabular}{ccc}
\includegraphics[width=\columnwidth,angle=0]{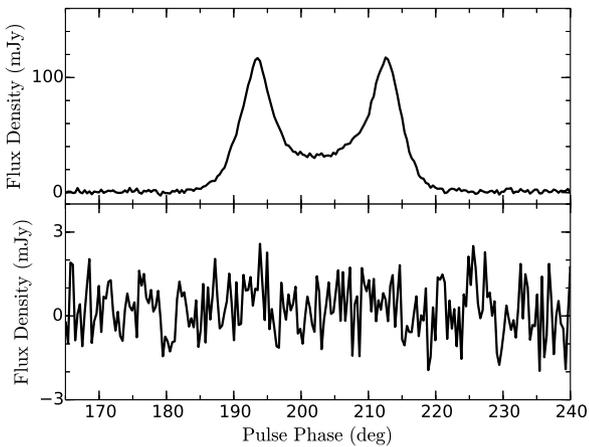}
\end{tabular}
\caption{Mean pulse profiles of null (lower) and burst pulses (upper)
  for the 2014-04-01 observation (see Table~\ref{tab:nf} for details).
}\label{fig:null_prf}
\end{figure}

\begin{figure}
\centering
\begin{tabular}{cc}
\includegraphics[width=\columnwidth,angle=0]{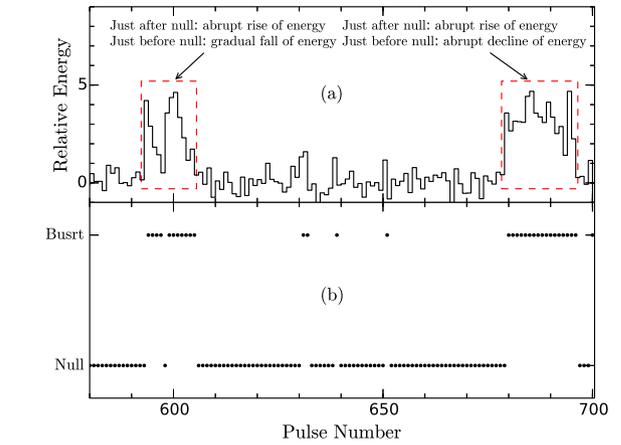} \\
\includegraphics[width=\columnwidth,angle=0]{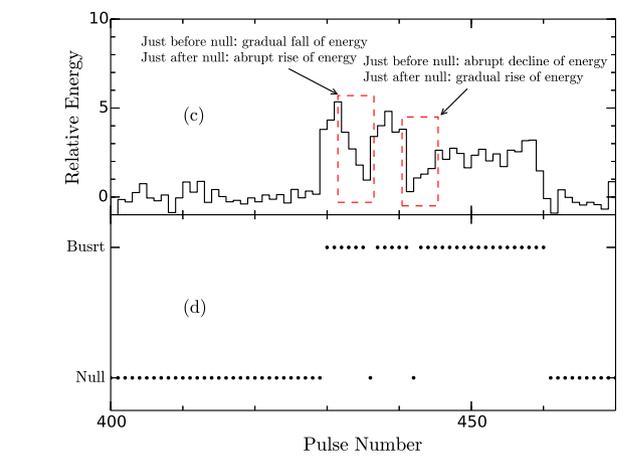}
\end{tabular}
\caption{Panels (a) and (c) show on-pulse energy vs. pulse number for two pulse
  sequences. Panels (b) and (d) show the identified burst or null states for
  each pulse.}
        \label{fig:null_trans}
\end{figure}

\begin{figure}
\begin{tabular}{cc}
\includegraphics[width=\columnwidth,angle=0]{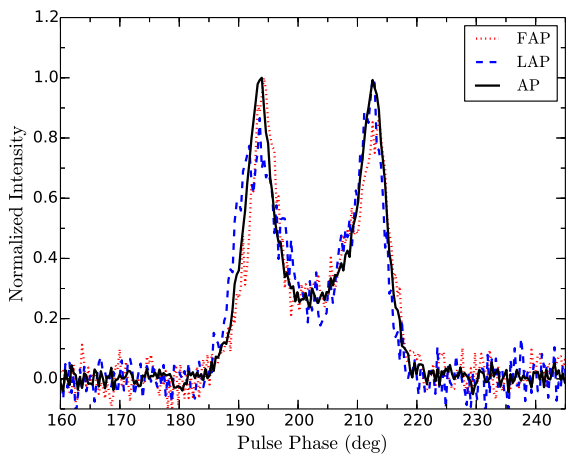}
\end{tabular}
\caption{Integrated profiles for the first active pulses (dotted line) and 
  the last active pulses (dashed line) of the burst states. The solid line
  shows the mean pulse profile of all burst pulses. All profiles
  are normalized by their respective peak intensities.
}\label{fig:FAP_LAP}
\end{figure}

\begin{figure}
\centering
\includegraphics[angle=0,width=\columnwidth]{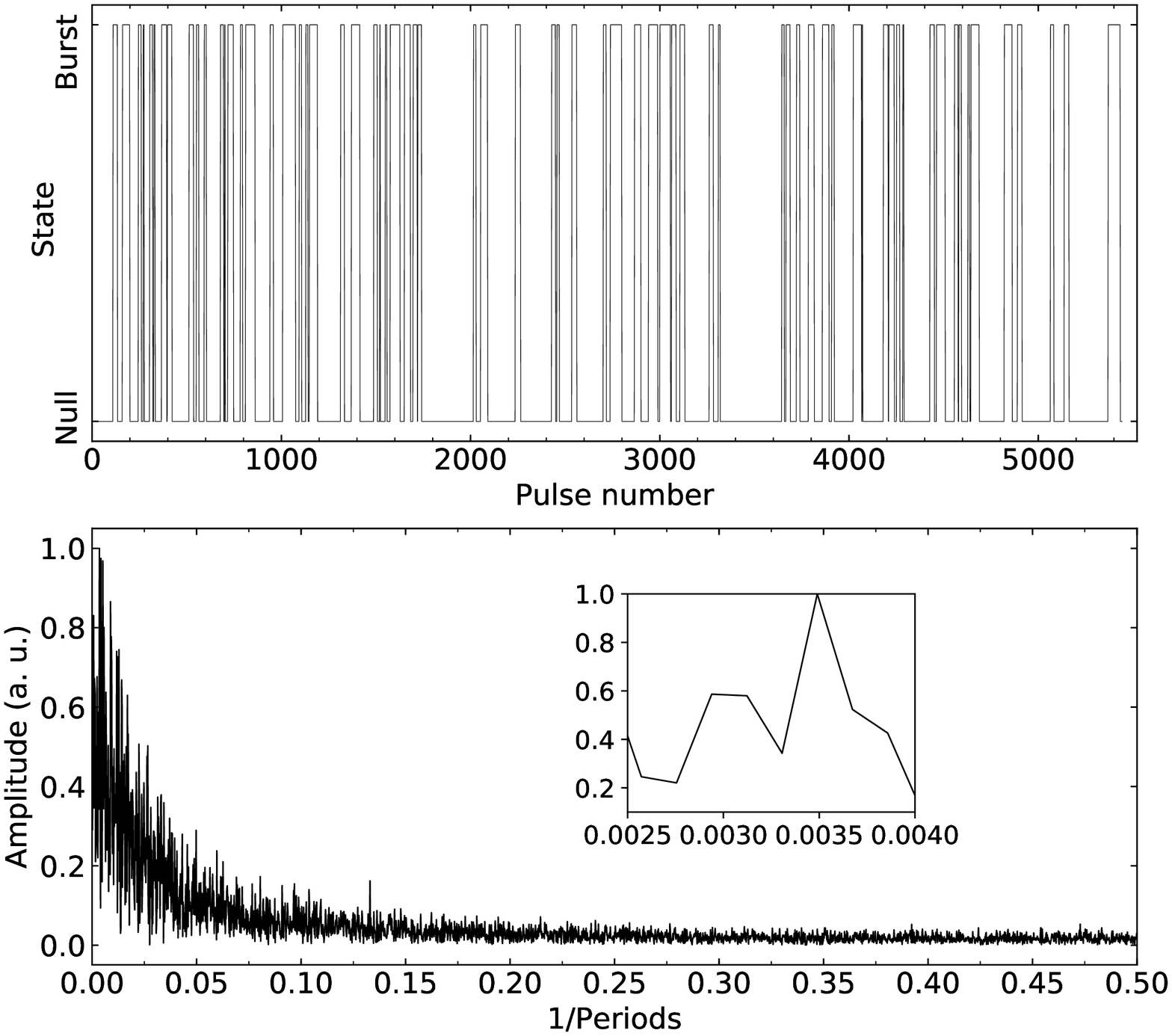}
\caption{Low-frequency quasi-periodic fluctuations in the pulse energy of
  PSR J1727$-$2739.
  The upper panel shows the identified emission states (null or burst)
  corresponding to each spin period for the 2014-04-03
    observation (see Table~\ref{tab:nf} for details).
  The lower panel shows the Fourier spectrum of the one-zero sequence
  (see the text for details) for the same observation. The inset
  plot exhibits the zoom in on the strongest spectral feature.
        \label{fig:period_null} }
    \end{figure}

The pulse energy variations with time of PSR J1727$-$2739 are presented in
Fig.~\ref{fig:energy_var} which shows many blocks of consecutive strong pulses
separated by frequent nulls.  The pulse energy distributions for the on-pulse and off-pulse
windows are shown in Fig.~\ref{fig:energy_dist}. We obtained the on-pulse energy for
each single pulse by summing the intensities of the pulse phase bins within the on-pulse
window of the mean pulse profile. The on-pulse window was defined as the total
longitude range over which significant
pulsed emission is seen. Similarly, the off-pulse energy was estimated using an off-pulse
window of the same length. In Fig.~\ref{fig:energy_dist}, the on-pulse energy histograms
show a clear bimodal distribution with two peaks, one at the zero energy and the other at
the mean pulse energy. The two distribution modes belong to null and burst states,
respectively.

To estimate the baseline noise level of each observation, we first normalized individual 
pulses with the peak intensity of the mean pulse profile in the corresponding day. Then the 
rms of the off-pulse region were estimated for all normalized single pulses.
The rms distributions for all five observations are given in Fig.~\ref{fig:rms_dist}
(see panels (a) -- (e)).
 The rms distribution of each of our observation 
can be fitted with a Gaussian function. The best fitted $\mu$ and $\sigma$ are the mean rms 
and its uncertainty, respectively, which are presented in Table~~\ref{tab:nf}.
For comparison, using the same method, we estimated the rms distribution for the
  observation of \citet{wwy+16} which is shown in panel
  (f) of  Fig.~\ref{fig:rms_dist}. Surprisingly, this distribution does not follow a
  Gaussian form. To explore the possible reason of the deviation from a Gaussian
  distribution, we compare the temporal variations of the off-pulse energy for observations
  of this study
  and \citet{wwy+16}, and the results are shown in Fig.~\ref{fig:eoff_var}. We can clearly
  see that the observation of \citet{wwy+16} shows random and significant fluctuations in
  off-pulse energy, which is probably due to the presence of strong RFI. We thefore
  suggest that the deviation from a Gaussian distribution for the rms distribution of
  \citet{wwy+16} probably arises from the relatively strong RFI.
The five distributions of our observations have slightly different mean values of the rms. 
To reduce the effect of different baseline rms noise levels between different observations,
we calculate the NF for every observation separately. Following earlier studies
\citep{rit76,wmj07,wwy+16}, we estimated the NF of each observation
by subtracting a scaled version of the off-pulse histogram from the on-pulse histogram
so that the sum of the difference counts in bins with $E<0$ was zero. The NF uncertainty is
then given by $\sqrt{n_{\rm p}}/N$, where $n_{\rm p}$ is the number of null pulses and 
$N$ is the total number of pulses \citep{wmj07}. The results for the five observations
are listed in Table~\ref{tab:nf}. The weighted average of NF of the five observations is
$66\%\pm1.4\%$ which is in good agreement with the previous work \citep{wwy+16}.

Identifying the burst and null  states is required for further investigation of the properties
of the burst and null pulses. 
We followed \citet{bgg10} to identify null pulses. First, we calculated the uncertainty
of the on-pulse energy $\sigma_{\rm ep}=\sqrt{n_{\rm on}}$ $\sigma_{\rm off}$ for each individual
pulse, where $n_{\rm on}$ is the number of on-pulse longitude bins and $\sigma_{\rm off}$ 
is the rms of the off-pulse region for individual pulses. Then, the
pulses with on-pulse energies 
lower than a threshold of $3\sigma_{\rm ep}$ are classified as null pulses and the others as
burst pulses.

The length distributions of identified burst and null are shown in
Fig.~\ref{fig:null_len}. Both burst and null lengths 
clearly cluster between two to five pulse periods, which means that the pulsar
switches frequently between burst and null states within several spin periods.
\citet{bac70} investigated the nulling phenomenon in four pulsars and he divided the nulls into
two types according to their duration: Type \uppercase\expandafter{\romannumeral1} nulls have a
width between three and ten pulses periods, whereas
Type \uppercase\expandafter{\romannumeral2} nulls
have a width of only one or two pulses. In this paper, we refer to Type
\uppercase\expandafter{\romannumeral1} nulls as long nulls and Type
\uppercase\expandafter{\romannumeral2} nulls as short nulls. Both types of nulls are observed
in PSR J1727$-$2739.

After the separation of null and burst states, the mean pulse profile obtained from
the null and the burst pulses are presented in Fig.~\ref{fig:null_prf}. In the lower
panel, the average profile of null pulses does not exhibit any significant 
emission component. This implies the absence of any detectable emission during the null state.

The transitional patterns between burst and null states help to understand the triggering and
transition mechanism. In PSR B0818$-$41, it was reported that the transitions from bursts to
nulls are gradual, while the transitions from nulls to bursts are abrupt \citep{bgg10}.
\citet{wwy+16} found an extra transitional pattern in PSR J1727$-$2739, in which the transitions
from bursts to nulls can be gradual or abrupt. However, they did not
report the transitional pattern
of short nulls. Two examples of the zoom-in view of pulse
energy variations with time to illustrate the transitional patterns are presented in
Fig.~\ref{fig:null_trans}. All transitional patterns reported by \citet{wwy+16} are shown in the
upper part of Fig.~\ref{fig:null_trans}. In the lower part, two short nulls show totally
different transitional patterns. 
The left short null shows that the transition from a burst to a null is gradual and the transitions
from a null to a burst is abrupt, whereas the right short null shows that the transition from a
burst to a null is abrupt and the transition from a null to a burst is gradual.

Differences in the shape of the average pulse profiles for the first active pulse (FAP)
after a null and the last active pulse (LAP) before a null have been reported in several pulsars
\citep[e.g.,][]{dchr86,gjw14}. We derived the average pulse profiles of the FAP and LAP for
PSR J1727$-$2739 and the result is shown in Fig.~\ref{fig:FAP_LAP}.
For the FAP profile, the peak intensities of the leading and trailing components are 1 and
  0.87, respectively, and the difference between the two peaks is 0.13.
For the LAP profile, the peak intensities of the leading and trailing components are 0.86 and
1, respectively, and the difference between the two peaks is 0.14.
For comparison, the rms values of baseline noise are 0.04 and 0.06 for the
FAP and LAP proflies, respectively.
Our results are in good agreement with those of \citet{wwy+16}.
The observed profile shape differences between the LAP and
FAP imply that the plasma conditions in the
magnetosphere may be different at the start and end of burst states.

\citet{bmm17} reported that the transitions between the burst and null states
of PSR J1727$-$2739 show a long periodicity of $206\pm33P_{1}$.
To investigate if the long periodicity exists in our data, we carried out
a similar analysis to \citet{gyy+17} for the three relatively longer observations
(2014-04-01, 2014-04-03 and 2014-10-14).
We set burst pulses as ones and null pulses as zeros, then a Fourier transform was
performed on this one-zero sequence and the results of the 2014-04-03 observation
are given in Fig.~\ref{fig:period_null}. 
Note that, following \citet{gyy+17}, short bursts
and short nulls that last for only one or two periods were ignored in
the one-zero sequence. 
A very clear quasi-periodic feature can be seen at very low frequency
in Fig.~\ref{fig:period_null}.
Results of the three longer observations were averaged and we finally obtained an average
periodicity of $243\pm65P_{1}$. This is consistent with the result of \citet{bmm17} within the
uncertainties.

\subsection{Subpulse drifting}
\label{sec:drift}

\begin{figure}
\includegraphics[width=\columnwidth,angle=0]{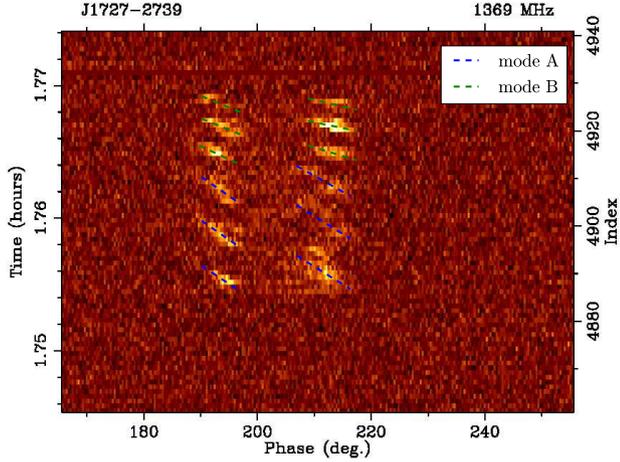}
\caption{A single-pulse stack showing an example of the rapid switching between 
different drift modes in the leading and trailing components.
 The drift bands are indicated by dashed lines. 
}\label{fig:drift_stack}
\end{figure}

\begin{figure}
\begin{tabular}{cc}
\includegraphics[width=0.9\columnwidth,angle=0]{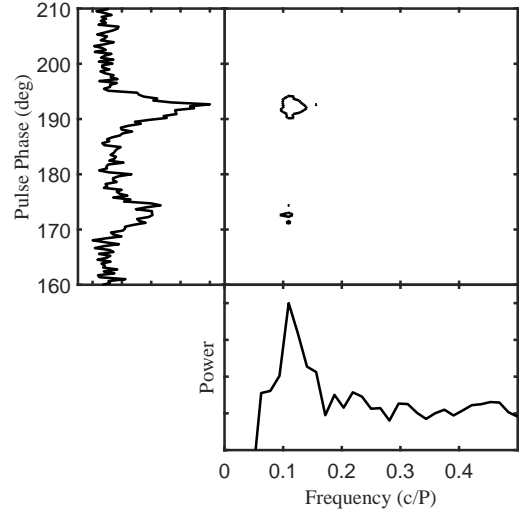}
\\\includegraphics[width=0.9\columnwidth,angle=0]{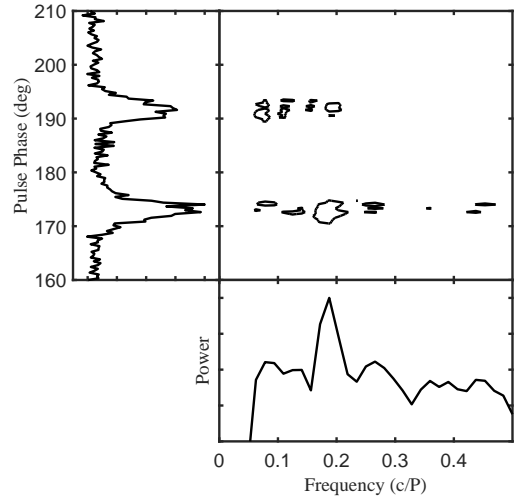}\\
\end{tabular}
\caption{Contour plots for the power spectrum of the flux density as a function of
  pulse phase for two drift sequences. The upper plot shows a 0.1 c/$P_{1}$
  periodicity, while the lower plot shows a 0.2 c/$P_{1}$ periodicity. The left
  side panels present the power integrated over frequency.
  The lower side panels give the power integrated over pulse phase.
 Power at frequencies lower than 0.05 c/$P_{1}$ has been set to zero.}\label{fig:paps}
\end{figure}

\begin{figure*}
\includegraphics[width=0.8\textwidth,angle=0]{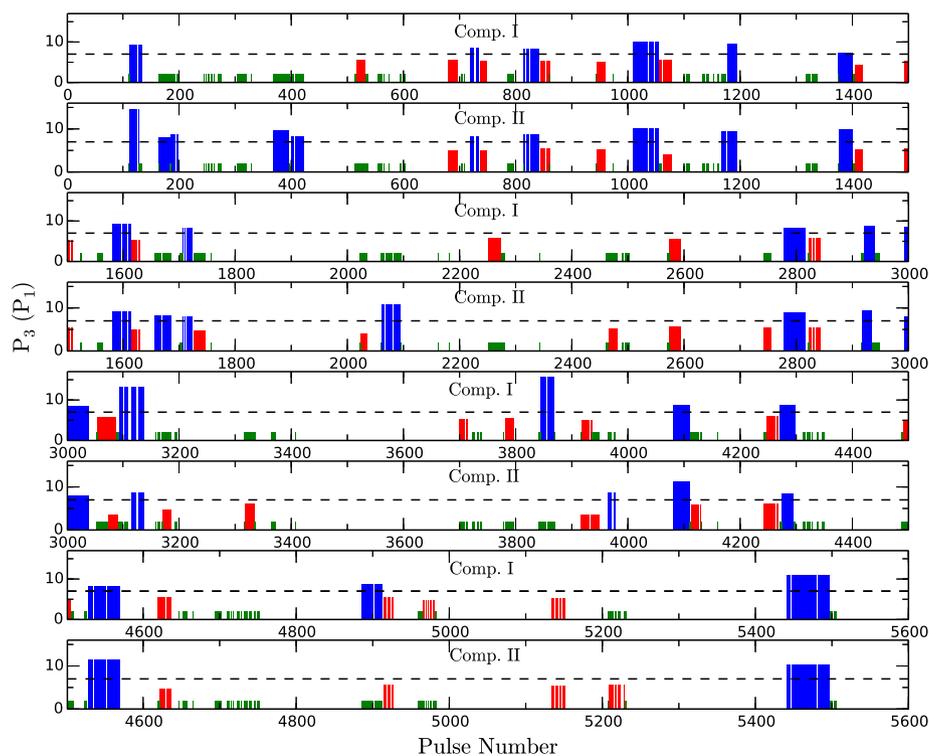}
\caption{Drift sequences showing observed $P_{3}$ values. Comp.
  {\rm \uppercase\expandafter{\romannumeral1}}
  and Comp. {\rm \uppercase\expandafter{\romannumeral2}}
  represent the leading and trailing components, respectively. 
  The modes A and B are plotted with blue and red bars, respectively.
  Green bars gives burst sequences showing no detection of subpulse drifting.
  The blank areas  correspond to nulls. $P_{3}=7$ is plotted with horizontal dashed-lines,  
which is the selected threshold for drift mode separation.
}\label{fig:drift_state}
\end{figure*}

\begin{figure*}
\begin{tabular}{cc}
\includegraphics[width=0.75\columnwidth,angle=-90.0]{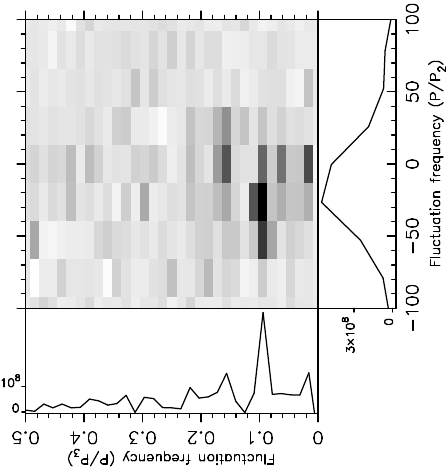}
&\includegraphics[width=0.75\columnwidth,angle=-90.0]{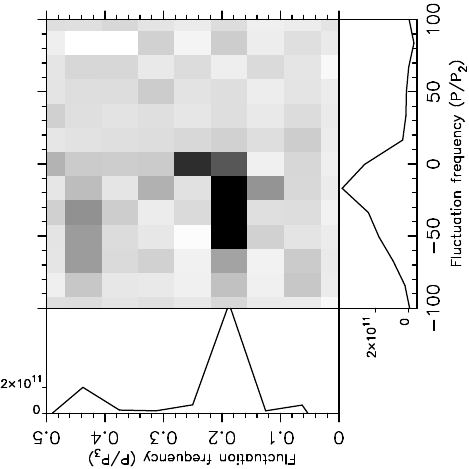}\\
\end{tabular}
\caption{The 2DFS with side panels showing horizontally (left) and vertically
  (bottom) integrated power for two drift sequences. The left plot shows the mode
  A of a sequece of 76 pulses with
  $P_{3} = 10.5\pm0.1~P_{1}$ and $P_{2} = 14\fdg3\pm1\fdg3$ in the leading
  component. The right plot shows the mode B of a sequence of 17 pulses with 
$P_{3} = 5.2\pm0.1~P_{1}$ and $P_{2} = 19\fdg6\pm2\fdg4$ in the trailing
component.
}\label{fig:2dfs}
\end{figure*}

\begin{figure}
\includegraphics[angle=0,width=\columnwidth]{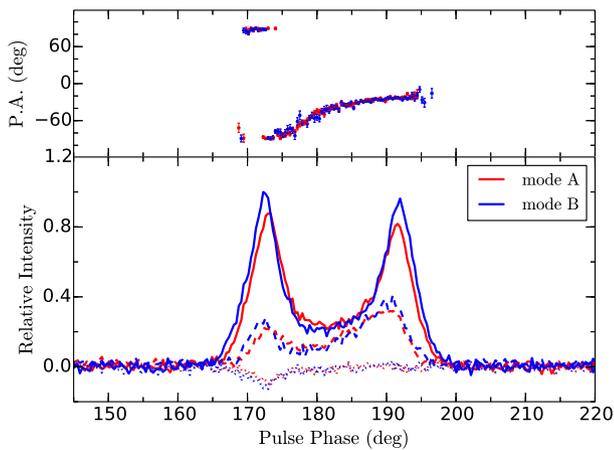}
\caption{Polarization profiles of different drift modes. For each
  mode, the lower
  panel shows the pulse profile for total intensity (solid line),
  linearly polarized intensity (dashed line),
  and circularly polarized intensity (dotted line).
  The upper panel gives the position angles of the linearly polarized emission.
}\label{fig:drift_poln}
\end{figure}

\begin{table*}
\centering
 \caption{Drift parameters for different drift modes. $\rm N_{\rm S}$ stands for number
 of sequences.}
\begin{tabular}{ccccccccccc}
\hline
Drift &$P_{\rm 3,l}$  &$P_{\rm 2,l}$  &$P_{\rm 3,t}$  &$P_{\rm 2,t}$ &$\Delta \phi_{\rm l}$  
&$\Delta \phi_{\rm t}$ &Number of &Number of   & $\rm W_{50}$  & $\rm W_{10}$  \\
mode   &$(P_{\rm 1})$    &$(\degr)$   &$(P_{\rm 1})$    &$(\degr)$  
 &  &  &sequences  &pulses  &$(\degr)$  &$(\degr)$  \\
  \hline
A  &$9.7\pm 0.4$  &$16.8\pm 2.8$  &$10.0\pm 0.3$   &$22.4\pm 3.7$ &$1.7\pm0.1$ &$2.2\pm0.1$ &103 &2264  &$23.8\pm0.3$  &$29.5\pm0.9$   \\
B  &$5.3\pm 0.2$  &$16.9\pm 3.0$  &$5.1\pm 0.1$   &$23.4\pm 4.1$  &$3.2\pm0.1$ &$4.6\pm0.1$ &97 &1605  &$24.5\pm0.4$  &$30.7\pm0.6$  \\
C  &$-$  &$-$ &$-$  &$-$   &$-$ &$-$   &$-$ &1083   &$24.5\pm0.4$  &$31.2\pm0.7$  \\
\hline
\end{tabular}
\label{tab:drift_mode}
%\end{minipage}

\end{table*} 

\begin{figure}
\includegraphics[angle=0,width=\columnwidth]{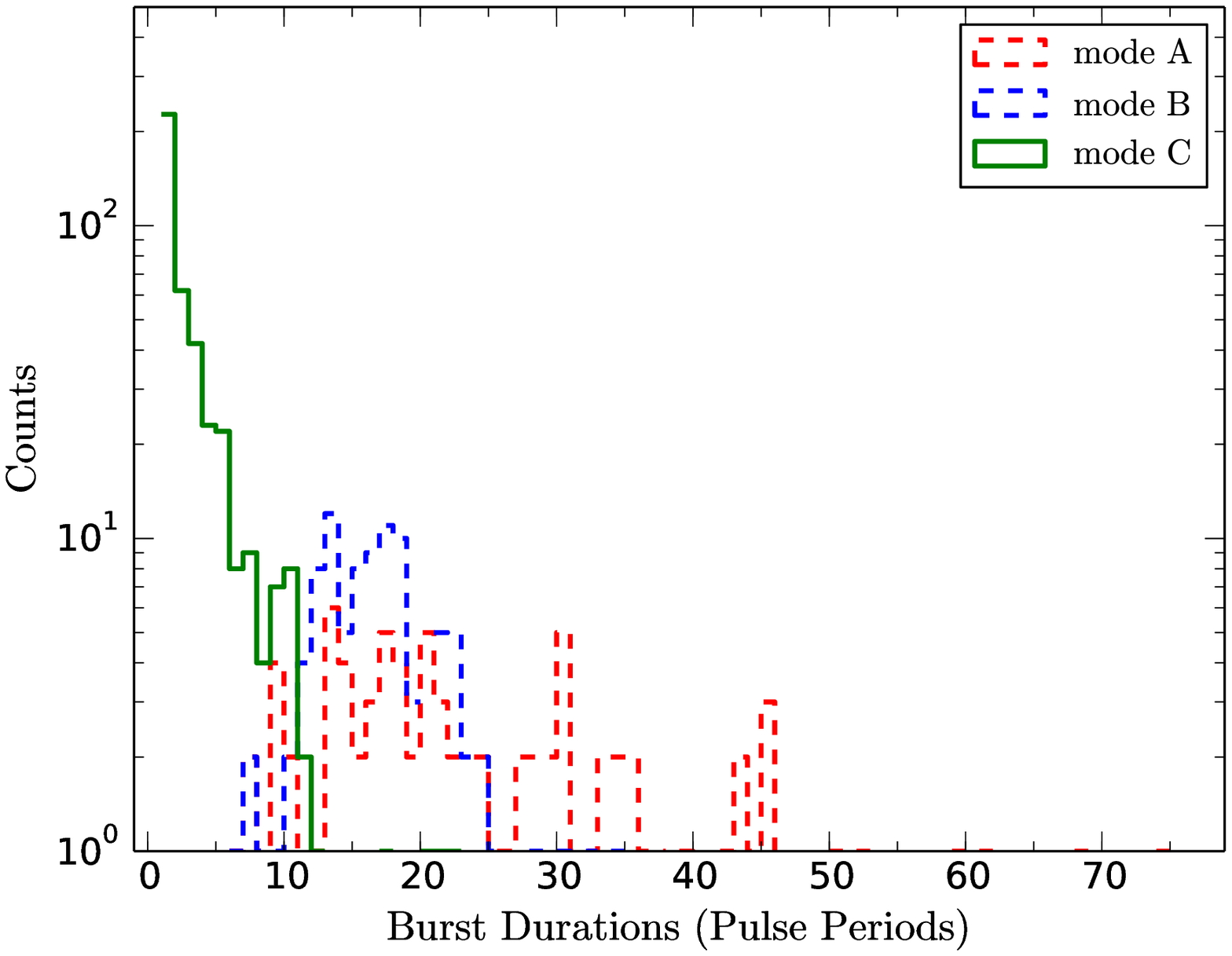}
\caption{Burst duration distributions of different drift modes.
}\label{fig:drift_dur}
\end{figure}

Based on the leading profile component, \citet{wwy+16} carried out a detailed
analysis of the subpulse drifting for PSR J1727$-$2739 at 1518 MHz. They
reported that this
pulsar shows two distinct drift modes, together with a non-drift mode.
The two drift modes were also detected at 610 MHz by \citet{bmm+21}.
\citet{wwy+16} also noticed the existence of an irregular drifting
in the trailing
component. However, no further detailed measurements had been performed,
probably due to the poor S/N of their data. 
Fig.~\ref{fig:drift_stack} shows a single pulse stack which exhibits clear subpulse 
drifting in both the leading and the trailing components.

In order to identify different drift modes, following earlier studies
\citep{smk05,wwy+16}, we used the phase-averaged power
spectrum (PAPS) to analyse our data. A candidate drift sequence was first 
selected visually. Then we computed the PAPS, which is the sum of the amplitudes
of the Discrete Fourier Transform (DFT) of each phase bin in a
selected pulse-phase
window, for this candidate
drift sequence. The start and
end of the sequence were then adjusted until the peak S/N of PAPS
(i.e., the peak value in the PAPS divided by the rms of the rest of the PAPS)
was maximized. The reciprocal of the frequency of the PAPS peak was calculated
as the drift periodicity $P_{3}$. As is shown in Fig.~\ref{fig:drift_stack},
drifting can be clearly seen in both profile components, we therefore
performed the PAPS analysis for both the leading and the trailing  components. 
Fig.~\ref{fig:paps} gives results of the PAPS analysis of two drift sequences
that belong to two different drift modes. 
The upper half of Fig.~\ref{fig:paps} shows the PAPS of a sequence of 31 pulses
with a peak around 0.1 c/$P_{1}$, which corresponds
to $P_{3}\sim$ 10$P_{1}$ for the trailing component. The lower half shows the PAPS of
a sequence of 30 pulses with a peak around 0.2 c/$P_{1}$,
which corresponds to $P_{3}\sim$ 5$P_{1}$ for the leading component.
 
Fig.~\ref{fig:drift_state} shows the $P_{3}$ values of both components for the
2014-04-01 observation. In agreement with \citep{wwy+16}, the $P_{3}$ values
of the leading component can be grouped into two distinct classes, viz.,
$P_{3}\sim$ 10 $P_{1}$ ( blue in
Fig.~\ref{fig:drift_state}) and $P_{3}\sim$ 5 $P_{1}$ 
( red in Fig.~\ref{fig:drift_state}).
The trailing component shows a very similar distribution of $P_{3}$ to the leading
component, which means that the trailing component also shows two drift modes
with similar periodicities to those of the leading component. This
indicates that the drift modes of both components have a common
origin.
By the PAPS technique, we can divide all burst pulse
sequences into different drift modes according to their $P_{3}$
values. Here, we defined pulse sequences with the leading
or trailing component showing $P_{3}\sim$10 $P_{1}$
as drift mode A, and those with the leading or trailing component 
showing $P_{3}\sim$ 5$P_{1}$ as drift mode B.
Pulse sequences without any detection of subpulse drifting in either
component were classified as drift mode C for convenience.
Based on Fig.~\ref{fig:drift_state}, we tried to find
evidence of interactions between nulling and subpulse drifting in
PSR J1727$-$2739. We found clues about links between
nulling and subpulse drifting. About 67\% of the drift sequences
of mode A follow behind nulls,
while more than 80\% of the drift sequences of mode B precede nulls.

After the drift mode separation,  we carried out an analysis of fluctuation spectra
for each drift sequence to investigate the
$P_{2}$ periodicities using the {\tt\string PSRSALSA} package.
Fig.~\ref{fig:2dfs} presents two examples of the resulted
two-dimensional fluctuation spectrum (2DFS, \citet{es02}).  
Table~\ref{tab:drift_mode}
shows a summary of the drift parameters
for different modes. Columns (2) and (3) give the values of $P_{3}$ and $P_{2}$,
respectively,
for the leading component. Columns (4) and (5) show the values of $P_{3}$ and $P_{2}$,
respectively, for the trailing component. Columns (6) and (7) present the drift rate 
$\bigtriangleup\phi$ for the leading and trailing components, respectively. Columns (8) 
and (9) give the number of sequences and the number of pulses, respectively. 
The last two columns show the pulse width at 50 per cent of the peak flux density
and pulse width at 10 per cent of the peak flux density, respectively.
Note that all values of $P_{3}$ and $P_{2}$ in Table~\ref{tab:drift_mode}
were averaged over drift sequences in corresponding drift modes.
Although both components have very similar $P_{3}$ in the same
drift mode, the $P_{2}$ we measured of the trailing component is
larger than that of the leading component by a factor of 1.3. Thus,
the drift rate of the trailing component is significantly larger than that
of the leading component in the same drift mode. This means that
the drift rate changes not only between drift modes but also between
profile components.
Like PSR B0031$-$07 where
$P_{2}$ for three different drift modes are the same \citep{htt70,smk05},
the $P_{2}$ of PSR J1727$-$2739 measured by us remains unchanged in different drift modes.
However, \citet{wwy+16} reported that the measured
$P_{2}$ for mode A is larger than that for mode B by a factor of 1.43.  
In our data, 43\% of burst pulses was in the drift mode A and 
30\% was in the drift mode B. By comparison, \citet{wwy+16} found that
the occurrence rate was 49\% for mode A and 29\% for mode B.

Fig.~\ref{fig:drift_poln} shows the polarization profiles
of different drift modes.
In general, there is no significant difference between
these polarization profiles. Their PA swings look identical. This implies that different
drift modes come from the same region in the magnetosphere.
The length distributions of different drift modes are shown in Fig.~\ref{fig:drift_dur}.
Our results are consistent with those of \citet{wwy+16}. The length of the
mode B clearly cluster between 5 to 35 pulse periods, whereas the
length of the drift mode A is more evenly distributed.

\section{Discussion and conclusions}\label{sec:discussion}

In this paper, we have conducted a detailed analysis of new single-pulse 
observations of PSR J1727$-$2739 with the Parkes 64-m radio telescope
at 1369 MHz. New results on the properties of nulling and subpulse
drifting for this pulsar have been reported.

In agreement with previous work \citep{wwy+16}, we estimate the NF to be
$66\%\pm1.4\%$ for
PSR J1727$-$2739 at 1369 MHz. To investigate whether the observed nulls are real
nulls or weak modes, we separated all the nulls and burst pulses. We did not find
any detectable emission in the average profile of null pulses. This suggests that
the observed nulls are true nulls and are not weak modes. The previously reported 
transitional patterns between bursts and nulls of PSR J1727$-$2739 were also observed
in this paper. Furthermore, we observed the transitional patterns between bursts and
short nulls that had not been published previously (Fig.~\ref{fig:null_trans}).
We confirm the existence of the long periodicity in the transitions between the burst
and null states which was discovered by \citet{bmm17} and our measured periodicity is
$243\pm65P_{1}$.
A study of periodic nulling and periodic amplitude modulations have been carried
out in a large number of pulsars by \citet{bmm20}.

The pattern of arrangement of nulls varies significantly from pulsar to pulsar. Even if
two pulsars have the same NF, their null lengths can be very different \citep{gjk12}.
Based on a detailed analysis of nulling for 36 pulsars, \citet{bmm17} concluded that
the most dominant nulls lasted for a short duration, less than five periods. Consistent
with such results, we found that the burst and null lengths of PSR J1727$-$2739
show similar distributions (Fig.~\ref{fig:null_len}) that cluster between two and
five pulse periods. However, \citet{wwy+16} presented significantly different distributions
of burst and null lengths. In \citet{wwy+16}, PSR J1727$-$2739 shows clustering at burst
length of around 20 pulse periods while the null length distribution seems to have a
favour of around 5$-$35 pule periods. Such differences probably result from the relatively
low S/N of the data of \citet{wwy+16}. In the case of lower S/N, it becomes difficult to identify
true null pulses from burst pulses. Differences in identification of true null and burst
pulses can cause large differences in the obtained null length and burst length distributions, 
specially near the short nulls and short bursts \citep{gaj17,cor13}.
Here, we classified pulses with on-pulse energy below the threshold 
$3\sigma_{\rm ep}$ as null pulses, whereas \citet{wwy+16} used a threshold of $5\sigma_{\rm ep}$.
We also tried to use $5\sigma_{\rm ep}$ as the threshold, but the derived average
profile of null pulses shows an obvious emission component. We therefore chose $3\sigma_{\rm ep}$,
which is more suitable for our data, as the threshold for identification of null and burst pulses.

Consistent with previously published results, we observe two distinct
drift modes in PSR J1727$-$2739: mode A (with $P_{3}\sim$ 10$P_{1}$)
and mode B (with $P_{3}\sim$ 5$P_{1}$). 
Compared with \citet{wwy+16},
we have some new findings about the drifting behaviour of PSR J1727$-$2739.
Apart from the previously known subpulse drifting in the leading component,
we found that the trailing component also shows subpulse drifting.
In a given drift mode, the drift periodicity $P_{3}$ is
constant across all profile components, but the measured $P_{2}$
values are quite different for both components. Contrary to the
previous results, we find that the periodicity
$P_{2}$ remains unchanged between drift modes A and B for each profile component.
Our results show that the drift rate of
the trailing component is larger than that of the leading component in
either drift mode. The polarization properties and the PA swings are very similar for
the two drift modes and this suggests that different drift modes originate from the same 
emission region. Such polarization behaviour remaining unchanged with the same
  PA swing in different emission modes has also been reported in PSRs J1822$-$2256
  \citep{bm18}, B0329+54 \citep{bmr19}, J2006$-$0807 \citep{bpm19} and J2321+6024
  \citep{rbmm21}.

The subpulse drifting phenomenon is traditionally explained by the carousel model \citep{rs75,vt12}, in
which discrete emission regions rotate around the magnetic axis. When the line of sight passes through
the emission region, drifting subpulses can be seen. The carousel model was successful in interpreting
subpulse drifting behaviour of many pulsars \citep{bggs07,rwb13,rr14,bil18}. In the original carousel
model, the drift rate remains constant, and thus it is difficult to explain the observed multiple drift
modes in a pulsar based on the original carousel model. There are
several models exploring the mode changing behaviour in pulsars, such
as reconfiguration of pulsar magnetosphere \citep{tim10}, multiple
magnetospheric states incorporating the apparent motion of the visible
point \citep{yue19}, the ion-proton pulsar polar cap model
\citep{jon20} and the partially screened gap
(PSG) model \citep{gmg03,smg15} in which the perturbations in the
magnetic field due to Hall and
thermal drift oscillations change the sparking configuration \citep{gbm+21}.
To explain the three distinct drift modes
observed in the conal triple pulsar PSR B1918+19
in the carousel framework, \citet{rwb13} explored the view that the observed drift bands result from
the first-order alias of a faster drift of subbeams equally spaced around the cones. They found that
in PSR B1918+19 the three drift modes have a common circulation period of $12P_{1}$ which is also the
periodicity of the quasi-periodic nulls. It is possible to extend the \citet{rwb13} model for the
drift modes of PSR J1727$-$2739 as well, but unlike PSR B1918+19, PSR J1727$-$2739 has a very
long null periodicity which is much longer than the periodicity
$P_{3}$ of the drift modes.

PSR J2321+6024 is a drifting pulsar showing three distinct
drift modes that have different $P_{3}$ values. 
  \citet{rbmm21} used the PSG model to explained the change of drift periodicity
of PSR J2321+6024, where a changing surface non-dipolar magnetic field structure is required.
  By simulation analysis, \citet{bmmg20}   showed that the different
 types of drifting phase behaviour can be explained by the PSG model
 using some assumptions of spark dynamics in a non-dipolar inner
 acceleration region (IAR). These studies imply that the PSG model
 is a very promising model for explaining the change of $P_{3}$ during
 drift mode changing and drifting phase behaviour. Thus, the PSG model
 is a very potential model to account for the drifting behaviour of
 PSR J1727$-$2739.

There is increasing evidence that nulling, mode changing and subpulse drifting are 
closely linked to each other. The presence of both nulling and
  subpulse drifting have been seen in a number of pulsars, such as PSRs B0031$-$07 
\citep{vj97,smk05,mbt+17}, J1822$-$2256 \citep{bm18}, B1944+17 
\citep{dchr86,kr10}, B2000+40 \citep{blk20}, J2006$-$0807 \citep{bpm19},
B2034+19 \citep{ran17}, J2321+6024 \citep{wf81,rbmm21} and B2303+30
\citep{rwr05}.
Some pulsars show changes in the drift rate after the null states \citep{vsrr03,jv04}.
Some pulsars exhibit memory of subpulse phases across nulls \citep{urwe78,gyy+17}. The interactions
between nulling and subpulse drifting suggest that  both nulls and subpulse drifting result from
intrinsic changes in the pulsar magnetosphere. 
Based on observations of 23 pulsars, \citet{wmj07} proposed that both nulling and mode
changing result from changes in the magnetospheric current distribution. Studies of nulling, mode
changing and subpulse drifting, and their interactions could provide a fundamental clue to the
underlying switching mechanism of changes between different magnetospheric states. 
We found that PSR J1727$-$2739 tends to start mode
A sequences after nulls and to start mode B sequences before
nulls.

PSR J1727$-$2739 falls in a rare category of pulsars showing
nulling along with drift mode changing,
which can provide unique opportunity to understand the links between these phenomena. 
Although we found evidence of interactions between nulling and
subpulse drifting in PSR J1727$-$2739, 
more comprehensive studies based on more sensitive, longer and
multifrequency observations for this kind
pulsars would certainly provide intriguing details to understand the
true nature and origin of
these phenomena.

\section*{Acknowledgements}

The Parkes radio telescope is part of the Australia Telescope National
Facility which is funded by the Commonwealth of Australia for
operation as a National Facility managed by CSIRO.
This work is supported by
National SKA Program of China No. 2020SKA0120200, 
the Joint Research
Fund in Astronomy under cooperative agreement between
the National Natural Science Foundation of China (NSFC)
and the Chinese Academy of Sciences (CAS) (No. U1831102,
U1731238), the NSFC project (No. 12041303, 12041304),
the open program of the Key Laboratory of Xinjiang Uygur
Autonomous Region No. 2020D04049 and the 
National Key Research and Development Program of China
(2016YFA0400804). This research is partly supported by
the Operation, Maintenance and Upgrading Fund for Astronomical 
Telescopes and Facility Instruments, budgeted from
the Ministry of Finance of China (MOF) and administrated
by the CAS.

%%%%%%%%%%%%%%%%%%%%%%%%%%%%%%%%%%%%%%%%%%%%%%%%%%
\section*{Data Availability}
 
The data underlying this article are available in the Parkes Pulsar Data
Archive at https://data.csiro.au, and can be accessed with the source name J1727$-$2739.



%\bibliographystyle{mnras}
%\bibliography{journals,modrefs,psrrefs,crossrefs} % if your bibtex file is called example.bib

\begin{thebibliography}{}
\makeatletter
\relax
\def\mn@urlcharsother{\let\do\@makeother \do\$\do\&\do\#\do\^\do\_\do\%\do\~}
\def\mn@doi{\begingroup\mn@urlcharsother \@ifnextchar [ {\mn@doi@}
  {\mn@doi@[]}}
\def\mn@doi@[#1]#2{\def\@tempa{#1}\ifx\@tempa\@empty \href
  {http://dx.doi.org/#2} {doi:#2}\else \href {http://dx.doi.org/#2} {#1}\fi
  \endgroup}
\def\mn@eprint#1#2{\mn@eprint@#1:#2::\@nil}
\def\mn@eprint@arXiv#1{\href {http://arxiv.org/abs/#1} {{\tt arXiv:#1}}}
\def\mn@eprint@dblp#1{\href {http://dblp.uni-trier.de/rec/bibtex/#1.xml}
  {dblp:#1}}
\def\mn@eprint@#1:#2:#3:#4\@nil{\def\@tempa {#1}\def\@tempb {#2}\def\@tempc
  {#3}\ifx \@tempc \@empty \let \@tempc \@tempb \let \@tempb \@tempa \fi \ifx
  \@tempb \@empty \def\@tempb {arXiv}\fi \@ifundefined
  {mn@eprint@\@tempb}{\@tempb:\@tempc}{\expandafter \expandafter \csname
  mn@eprint@\@tempb\endcsname \expandafter{\@tempc}}}

\bibitem[\protect\citeauthoryear{{Backer}}{{Backer}}{1970a}]{bac70}
{Backer} D.~C.,  1970a, \mn@doi [Nature] {10.1038/228042a0}, \href
  {https://ui.adsabs.harvard.edu/abs/1970Natur.228...42B} {228, 42}

\bibitem[\protect\citeauthoryear{{Backer}}{{Backer}}{1970b}]{bac70a}
{Backer} D.~C.,  1970b, \mn@doi [Nature] {10.1038/2281297a0}, \href
  {https://ui.adsabs.harvard.edu/abs/1970Natur.228.1297B} {228, 1297}

\bibitem[\protect\citeauthoryear{{Basu} \& {Mitra}}{{Basu} \&
  {Mitra}}{2018}]{bm18}
{Basu} R.,  {Mitra} D.,  2018, \mn@doi [MNRAS] {10.1093/mnras/sty297}, \href
  {https://ui.adsabs.harvard.edu/abs/2018MNRAS.476.1345B} {476, 1345}

\bibitem[\protect\citeauthoryear{{Basu} \& {Mitra}}{{Basu} \&
  {Mitra}}{2019}]{bm19}
{Basu} R.,  {Mitra} D.,  2019, \mn@doi [MNRAS] {10.1093/mnras/stz1590}, \href
  {https://ui.adsabs.harvard.edu/abs/2019MNRAS.487.4536B} {487, 4536}

\bibitem[\protect\citeauthoryear{{Basu}, {Mitra}, {Melikidze}, {Maciesiak},
  {Skrzypczak}  \& {Szary}}{{Basu} et~al.}{2016}]{bmm+16}
{Basu} R.,  {Mitra} D.,  {Melikidze} G.~I.,  {Maciesiak} K.,  {Skrzypczak} A.,
   {Szary} A.,  2016, \mn@doi [ApJ] {10.3847/1538-4357/833/1/29}, \href
  {http://adsabs.harvard.edu/abs/2016ApJ...833...29B} {833, 29}

\bibitem[\protect\citeauthoryear{{Basu}, {Mitra}  \& {Melikidze}}{{Basu}
  et~al.}{2017}]{bmm17}
{Basu} R.,  {Mitra} D.,   {Melikidze} G.~I.,  2017, \mn@doi [ApJ]
  {10.3847/1538-4357/aa862d}, \href
  {http://adsabs.harvard.edu/abs/2017ApJ...846..109B} {846, 109}

\bibitem[\protect\citeauthoryear{{Basu}, {Mitra}, {Melikidze}  \&
  {Skrzypczak}}{{Basu} et~al.}{2019a}]{bmm+19}
{Basu} R.,  {Mitra} D.,  {Melikidze} G.~I.,   {Skrzypczak} A.,  2019a, \mn@doi
  [MNRAS] {10.1093/mnras/sty2846}, \href
  {https://ui.adsabs.harvard.edu/abs/2019MNRAS.482.3757B} {482, 3757}

\bibitem[\protect\citeauthoryear{{Basu}, {Paul}  \& {Mitra}}{{Basu}
  et~al.}{2019b}]{bpm19}
{Basu} R.,  {Paul} A.,   {Mitra} D.,  2019b, \mn@doi [MNRAS]
  {10.1093/mnras/stz1225}, \href
  {https://ui.adsabs.harvard.edu/abs/2019MNRAS.486.5216B} {486, 5216}

\bibitem[\protect\citeauthoryear{{Basu}, {Mitra}  \& {Melikidze}}{{Basu}
  et~al.}{2020a}]{bmmg20}
{Basu} R.,  {Mitra} D.,   {Melikidze} G.~I.,  2020a, \mn@doi [MNRAS]
  {10.1093/mnras/staa1574}, \href
  {https://ui.adsabs.harvard.edu/abs/2020MNRAS.496..465B} {496, 465}

\bibitem[\protect\citeauthoryear{{Basu}, {Lewandowski}  \& {Kijak}}{{Basu}
  et~al.}{2020b}]{blk20}
{Basu} R.,  {Lewandowski} W.,   {Kijak} J.,  2020b, \mn@doi [MNRAS]
  {10.1093/mnras/staa2398}, \href
  {https://ui.adsabs.harvard.edu/abs/2020MNRAS.499..906B} {499, 906}

\bibitem[\protect\citeauthoryear{{Basu}, {Mitra}  \& {Melikidze}}{{Basu}
  et~al.}{2020c}]{bmm20}
{Basu} R.,  {Mitra} D.,   {Melikidze} G.~I.,  2020c, \mn@doi [ApJ]
  {10.3847/1538-4357/ab63c9}, \href
  {https://ui.adsabs.harvard.edu/abs/2020ApJ...889..133B} {889, 133}

\bibitem[\protect\citeauthoryear{{Basu}, {Mitra}  \& {Melikidze}}{{Basu}
  et~al.}{2021}]{bmm+21}
{Basu} R.,  {Mitra} D.,   {Melikidze} G.~I.,  2021, \mn@doi [\apj]
  {10.3847/1538-4357/ac0828}, \href
  {https://ui.adsabs.harvard.edu/abs/2021ApJ...917...48B} {917, 48}

\bibitem[\protect\citeauthoryear{{Bhattacharyya}, {Gupta}, {Gil}  \&
  {Sendyk}}{{Bhattacharyya} et~al.}{2007}]{bggs07}
{Bhattacharyya} B.,  {Gupta} Y.,  {Gil} J.,   {Sendyk} M.,  2007, \mn@doi
  [MNRAS] {10.1111/j.1745-3933.2007.00293.x}, \href
  {https://ui.adsabs.harvard.edu/abs/2007MNRAS.377L..10B} {377, L10}

\bibitem[\protect\citeauthoryear{{Bhattacharyya}, {Gupta}  \&
  {Gil}}{{Bhattacharyya} et~al.}{2010}]{bgg10}
{Bhattacharyya} B.,  {Gupta} Y.,   {Gil} J.,  2010, \mn@doi [MNRAS]
  {10.1111/j.1365-2966.2010.17116.x}, \href
  {http://adsabs.harvard.edu/abs/2010MNRAS.408..407B} {408, 407}

\bibitem[\protect\citeauthoryear{{Bilous}}{{Bilous}}{2018}]{bil18}
{Bilous} A.~V.,  2018, \mn@doi [A\&A] {10.1051/0004-6361/201732106}, \href
  {https://ui.adsabs.harvard.edu/abs/2018A&A...616A.119B} {616, A119}

\bibitem[\protect\citeauthoryear{{Brinkman}, {Mitra}  \& {Rankin}}{{Brinkman}
  et~al.}{2019}]{bmr19}
{Brinkman} C.,  {Mitra} D.,   {Rankin} J.,  2019, \mn@doi [\mnras]
  {10.1093/mnras/stz020}, \href
  {https://ui.adsabs.harvard.edu/abs/2019MNRAS.484.2725B} {484, 2725}

\bibitem[\protect\citeauthoryear{{Cordes}}{{Cordes}}{2013}]{cor13}
{Cordes} J.~M.,  2013, \mn@doi [ApJ] {10.1088/0004-637X/775/1/47}, \href
  {https://ui.adsabs.harvard.edu/abs/2013ApJ...775...47C} {775, 47}

\bibitem[\protect\citeauthoryear{Deich, Cordes, Hankins  \& Rankin}{Deich
  et~al.}{1986}]{dchr86}
Deich W. T.~S.,  Cordes J.~M.,  Hankins T.~H.,   Rankin J.~M.,  1986, \mn@doi
  [ApJ] {10.1086/163831}, \href
  {https://ui.adsabs.harvard.edu/abs/1986ApJ...300..540D} {300, 540}

\bibitem[\protect\citeauthoryear{Drake \& Craft}{Drake \& Craft}{1968}]{dc68}
Drake F.~D.,  Craft H.~D.,  1968, \mn@doi [Nature] {10.1038/220231a0}, \href
  {https://ui.adsabs.harvard.edu/abs/1968Natur.220..231D} {220, 231}

\bibitem[\protect\citeauthoryear{{Edwards} \& {Stappers}}{{Edwards} \&
  {Stappers}}{2002}]{es02}
{Edwards} R.~T.,  {Stappers} B.~W.,  2002, A\&A, 393, 733

\bibitem[\protect\citeauthoryear{{Force} \& {Rankin}}{{Force} \&
  {Rankin}}{2010}]{fr10}
{Force} M.~M.,  {Rankin} J.~M.,  2010, \mn@doi [MNRAS]
  {10.1111/j.1365-2966.2010.16703.x}, \href
  {https://ui.adsabs.harvard.edu/abs/2010MNRAS.406..237F} {406, 237}

\bibitem[\protect\citeauthoryear{{Gajjar}}{{Gajjar}}{2017}]{gaj17}
{Gajjar} V.,  2017, arXiv e-prints, \href
  {https://ui.adsabs.harvard.edu/abs/2017arXiv170605407G} {p. arXiv:1706.05407}

\bibitem[\protect\citeauthoryear{{Gajjar}, {Joshi}  \& {Kramer}}{{Gajjar}
  et~al.}{2012}]{gjk12}
{Gajjar} V.,  {Joshi} B.~C.,   {Kramer} M.,  2012, \mn@doi [MNRAS]
  {10.1111/j.1365-2966.2012.21296.x}, \href
  {http://adsabs.harvard.edu/abs/2012MNRAS.424.1197G} {424, 1197}

\bibitem[\protect\citeauthoryear{{Gajjar}, {Joshi}  \& {Wright}}{{Gajjar}
  et~al.}{2014}]{gjw14}
{Gajjar} V.,  {Joshi} B.~C.,   {Wright} G.,  2014, \mn@doi [MNRAS]
  {10.1093/mnras/stt2389}, \href
  {http://adsabs.harvard.edu/abs/2014MNRAS.439..221G} {439, 221}

\bibitem[\protect\citeauthoryear{{Gajjar}, {Yuan}, {Yuen}, {Wen}, {Liu}  \&
  {Wang}}{{Gajjar} et~al.}{2017}]{gyy+17}
{Gajjar} V.,  {Yuan} J.~P.,  {Yuen} R.,  {Wen} Z.~G.,  {Liu} Z.~Y.,   {Wang}
  N.,  2017, \mn@doi [ApJ] {10.3847/1538-4357/aa96ac}, \href
  {http://adsabs.harvard.edu/abs/2017ApJ...850..173G} {850, 173}

\bibitem[\protect\citeauthoryear{{Geppert}, {Basu}, {Mitra}, {Melikidze}  \&
  {Szkudlarek}}{{Geppert} et~al.}{2021}]{gbm+21}
{Geppert} U.,  {Basu} R.,  {Mitra} D.,  {Melikidze} G.~I.,   {Szkudlarek} M.,
  2021, \mn@doi [MNRAS] {10.1093/mnras/stab1134}, \href
  {https://ui.adsabs.harvard.edu/abs/2021MNRAS.504.5741G} {504, 5741}

\bibitem[\protect\citeauthoryear{{Gil}, {Melikidze}  \& {Geppert}}{{Gil}
  et~al.}{2003}]{gmg03}
{Gil} J.,  {Melikidze} G.~I.,   {Geppert} U.,  2003, \mn@doi [A\&A]
  {10.1051/0004-6361:20030854}, \href
  {https://ui.adsabs.harvard.edu/abs/2003A&A...407..315G} {407, 315}

\bibitem[\protect\citeauthoryear{{Herfindal} \& {Rankin}}{{Herfindal} \&
  {Rankin}}{2007}]{hr07}
{Herfindal} J.~L.,  {Rankin} J.~M.,  2007, \mn@doi [MNRAS]
  {10.1111/j.1365-2966.2007.12089.x}, \href
  {http://adsabs.harvard.edu/abs/2007MNRAS.380..430H} {380, 430}

\bibitem[\protect\citeauthoryear{{Herfindal} \& {Rankin}}{{Herfindal} \&
  {Rankin}}{2009}]{hr09}
{Herfindal} J.~L.,  {Rankin} J.~M.,  2009, \mn@doi [MNRAS]
  {10.1111/j.1365-2966.2008.14119.x}, \href
  {http://adsabs.harvard.edu/abs/2009MNRAS.393.1391H} {393, 1391}

\bibitem[\protect\citeauthoryear{{Hobbs} et~al.,}{{Hobbs}
  et~al.}{2004}]{hfs+04}
{Hobbs} G.,  et~al., 2004, \mn@doi [MNRAS] {10.1111/j.1365-2966.2004.08042.x},
  \href {https://ui.adsabs.harvard.edu/abs/2004MNRAS.352.1439H} {352, 1439}

\bibitem[\protect\citeauthoryear{{Hobbs} et~al.,}{{Hobbs}
  et~al.}{2011}]{hmm+11}
{Hobbs} G.,  et~al., 2011, \mn@doi [PASA] {10.1071/AS11016}, \href
  {http://adsabs.harvard.edu/abs/2011PASA...28..202H} {28, 202}

\bibitem[\protect\citeauthoryear{{Hotan}, {van Straten}  \&
  {Manchester}}{{Hotan} et~al.}{2004}]{hvm04}
{Hotan} A.~W.,  {van Straten} W.,   {Manchester} R.~N.,  2004, \mn@doi [PASA]
  {10.1071/AS04022}, \href
  {https://ui.adsabs.harvard.edu/abs/2004PASA...21..302H} {21, 302}

\bibitem[\protect\citeauthoryear{Huguenin, Taylor  \& Troland}{Huguenin
  et~al.}{1970}]{htt70}
Huguenin G.~R.,  Taylor J.~H.,   Troland T.~H.,  1970, ApJ, 162, 727

\bibitem[\protect\citeauthoryear{{Janssen} \& {van Leeuwen}}{{Janssen} \& {van
  Leeuwen}}{2004}]{jv04}
{Janssen} G.~H.,  {van Leeuwen} J.,  2004, \mn@doi [A\&A]
  {10.1051/0004-6361:20041062}, 425, 255

\bibitem[\protect\citeauthoryear{{Jones}}{{Jones}}{2020}]{jon20}
{Jones} P.~B.,  2020, \mn@doi [MNRAS] {10.1093/mnras/staa247}, \href
  {https://ui.adsabs.harvard.edu/abs/2020MNRAS.492.5987J} {492, 5987}

\bibitem[\protect\citeauthoryear{{Kloumann} \& {Rankin}}{{Kloumann} \&
  {Rankin}}{2010}]{kr10}
{Kloumann} I.~M.,  {Rankin} J.~M.,  2010, \mn@doi [MNRAS]
  {10.1111/j.1365-2966.2010.17114.x}, \href
  {http://adsabs.harvard.edu/abs/2010MNRAS.408...40K} {408, 40}

\bibitem[\protect\citeauthoryear{{Manchester}, {Hobbs}, {Teoh}  \&
  {Hobbs}}{{Manchester} et~al.}{2005}]{mhth05}
{Manchester} R.~N.,  {Hobbs} G.~B.,  {Teoh} A.,   {Hobbs} M.,  2005, \mn@doi
  [AJ] {10.1086/428488}, \href
  {https://ui.adsabs.harvard.edu/abs/2005AJ....129.1993M} {129, 1993}

\bibitem[\protect\citeauthoryear{{Manchester} et~al.,}{{Manchester}
  et~al.}{2013}]{mhb+13}
{Manchester} R.~N.,  et~al., 2013, \mn@doi [PASA] {10.1017/pasa.2012.017},
  \href {http://adsabs.harvard.edu/abs/2013PASA...30...17M} {30, e017}

\bibitem[\protect\citeauthoryear{{McSweeney}, {Bhat}, {Tremblay}, {Deshpande}
  \& {Ord}}{{McSweeney} et~al.}{2017}]{mbt+17}
{McSweeney} S.~J.,  {Bhat} N.~D.~R.,  {Tremblay} S.~E.,  {Deshpande} A.~A.,
  {Ord} S.~M.,  2017, \mn@doi [ApJ] {10.3847/1538-4357/aa5c35}, \href
  {https://ui.adsabs.harvard.edu/abs/2017ApJ...836..224M} {836, 224}

\bibitem[\protect\citeauthoryear{{Mitra} \& {Rankin}}{{Mitra} \&
  {Rankin}}{2017}]{mr17}
{Mitra} D.,  {Rankin} J.,  2017, \mn@doi [MNRAS] {10.1093/mnras/stx814}, \href
  {http://adsabs.harvard.edu/abs/2017MNRAS.468.4601M} {468, 4601}

\bibitem[\protect\citeauthoryear{{Rahaman}, {Basu}, {Mitra}  \&
  {Melikidze}}{{Rahaman} et~al.}{2021}]{rbmm21}
{Rahaman} S. k.~M.,  {Basu} R.,  {Mitra} D.,   {Melikidze} G.~I.,  2021,
  \mn@doi [MNRAS] {10.1093/mnras/staa3518}, \href
  {https://ui.adsabs.harvard.edu/abs/2021MNRAS.500.4139R} {500, 4139}

\bibitem[\protect\citeauthoryear{Rankin}{Rankin}{1986}]{ran86}
Rankin J.~M.,  1986, \mn@doi [ApJ] {10.1086/163955}, \href
  {https://ui.adsabs.harvard.edu/abs/1986ApJ...301..901R} {301, 901}

\bibitem[\protect\citeauthoryear{{Rankin}}{{Rankin}}{2017}]{ran17}
{Rankin} J.~M.,  2017, \mn@doi [JA\&A] {10.1007/s12036-017-9462-9}, \href
  {https://ui.adsabs.harvard.edu/abs/2017JApA...38...53R} {38, 53}

\bibitem[\protect\citeauthoryear{{Rankin} \& {Rosen}}{{Rankin} \&
  {Rosen}}{2014}]{rr14}
{Rankin} J.,  {Rosen} R.,  2014, \mn@doi [MNRAS] {10.1093/mnras/stu237}, \href
  {https://ui.adsabs.harvard.edu/abs/2014MNRAS.439.3860R} {439, 3860}

\bibitem[\protect\citeauthoryear{{Rankin} \& {Wright}}{{Rankin} \&
  {Wright}}{2008}]{rw08}
{Rankin} J.~M.,  {Wright} G.~A.~E.,  2008, \mn@doi [MNRAS]
  {10.1111/j.1365-2966.2008.13001.x}, \href
  {http://adsabs.harvard.edu/abs/2008MNRAS.385.1923R} {385, 1923}

\bibitem[\protect\citeauthoryear{{Rankin}, {Wright}  \& {Brown}}{{Rankin}
  et~al.}{2013}]{rwb13}
{Rankin} J.~M.,  {Wright} G.~A.~E.,   {Brown} A.~M.,  2013, \mn@doi [MNRAS]
  {10.1093/mnras/stt739}, \href
  {http://adsabs.harvard.edu/abs/2013MNRAS.433..445R} {433, 445}

\bibitem[\protect\citeauthoryear{{Redman}, {Wright}  \& {Rankin}}{{Redman}
  et~al.}{2005}]{rwr05}
{Redman} S.~L.,  {Wright} G.~A.~E.,   {Rankin} J.~M.,  2005, \mn@doi [MNRAS]
  {10.1111/j.1365-2966.2005.08672.x}, \href
  {https://ui.adsabs.harvard.edu/abs/2005MNRAS.357..859R} {357, 859}

\bibitem[\protect\citeauthoryear{Ritchings}{Ritchings}{1976}]{rit76}
Ritchings R.~T.,  1976, \mn@doi [MNRAS] {10.1093/mnras/176.2.249}, \href
  {https://ui.adsabs.harvard.edu/abs/1976MNRAS.176..249R} {176, 249}

\bibitem[\protect\citeauthoryear{Ruderman \& Sutherland}{Ruderman \&
  Sutherland}{1975}]{rs75}
Ruderman M.~A.,  Sutherland P.~G.,  1975, ApJ, 196, 51

\bibitem[\protect\citeauthoryear{{Smits}, {Mitra}  \& {Kuijpers}}{{Smits}
  et~al.}{2005}]{smk05}
{Smits} J.~M.,  {Mitra} D.,   {Kuijpers} J.,  2005, \mn@doi [A\&A]
  {10.1051/0004-6361:20041626}, \href
  {https://ui.adsabs.harvard.edu/abs/2005A&A...440..683S} {440, 683}

\bibitem[\protect\citeauthoryear{Staveley-Smith et~al.,}{Staveley-Smith
  et~al.}{1996}]{swb+96}
Staveley-Smith L.,  et~al., 1996, PASA, \href
  {https://ui.adsabs.harvard.edu/abs/1996PASA...13..243S} {13, 243}

\bibitem[\protect\citeauthoryear{{Szary}, {Melikidze}  \& {Gil}}{{Szary}
  et~al.}{2015}]{smg15}
{Szary} A.,  {Melikidze} G.~I.,   {Gil} J.,  2015, \mn@doi [MNRAS]
  {10.1093/mnras/stu2622}, \href
  {https://ui.adsabs.harvard.edu/abs/2015MNRAS.447.2295S} {447, 2295}

\bibitem[\protect\citeauthoryear{{Timokhin}}{{Timokhin}}{2010}]{tim10}
{Timokhin} A.~N.,  2010, \mn@doi [MNRAS] {10.1111/j.1745-3933.2010.00924.x},
  \href {https://ui.adsabs.harvard.edu/abs/2010MNRAS.408L..41T} {408, L41}

\bibitem[\protect\citeauthoryear{Unwin, Readhead, Wilkinson  \& Ewing}{Unwin
  et~al.}{1978}]{urwe78}
Unwin S.~C.,  Readhead A. C.~S.,  Wilkinson P.~N.,   Ewing M.~S.,  1978, MNRAS,
182, 711

\bibitem[\protect\citeauthoryear{{van Leeuwen} \& {Timokhin}}{{van Leeuwen} \&
  {Timokhin}}{2012}]{vt12}
{van Leeuwen} J.,  {Timokhin} A.~N.,  2012, \mn@doi [ApJ]
  {10.1088/0004-637X/752/2/155}, \href
  {https://ui.adsabs.harvard.edu/abs/2012ApJ...752..155V} {752, 155}

\bibitem[\protect\citeauthoryear{{van Leeuwen}, {Stappers}, {Ramachandran}  \&
  {Rankin}}{{van Leeuwen} et~al.}{2003}]{vsrr03}
{van Leeuwen} A.~G.~J.,  {Stappers} B.~W.,  {Ramachandran} R.,   {Rankin}
  J.~M.,  2003, \mn@doi [A\&A] {10.1051/0004-6361:20021630}, \href
  {https://ui.adsabs.harvard.edu/abs/2003A&A...399..223V} {399, 223}

\bibitem[\protect\citeauthoryear{{van Straten} \& {Bailes}}{{van Straten} \&
  {Bailes}}{2011}]{vb11}
{van Straten} W.,  {Bailes} M.,  2011, \mn@doi [PASA] {10.1071/AS10021}, \href
  {http://adsabs.harvard.edu/abs/2011PASA...28....1V} {28, 1}

\bibitem[\protect\citeauthoryear{{Vivekanand} \& {Joshi}}{{Vivekanand} \&
  {Joshi}}{1997}]{vj97}
{Vivekanand} M.,  {Joshi} B.~C.,  1997, \mn@doi [ApJ] {10.1086/303690}, \href
  {https://ui.adsabs.harvard.edu/abs/1997ApJ...477..431V} {477, 431}

\bibitem[\protect\citeauthoryear{{Wang}, {Manchester}  \& {Johnston}}{{Wang}
  et~al.}{2007}]{wmj07}
{Wang} N.,  {Manchester} R.~N.,   {Johnston} S.,  2007, \mn@doi [MNRAS]
  {10.1111/j.1365-2966.2007.11703.x}, \href
  {https://ui.adsabs.harvard.edu/abs/2007MNRAS.377.1383W} {377, 1383}

\bibitem[\protect\citeauthoryear{{Wang} et~al.,}{{Wang} et~al.}{2020}]{whh+20}
{Wang} P.~F.,  et~al., 2020, \mn@doi [A\&A] {10.1051/0004-6361/202038867},
  \href {https://ui.adsabs.harvard.edu/abs/2020A&A...644A..73W} {644, A73}

\bibitem[\protect\citeauthoryear{{Weltevrede}}{{Weltevrede}}{2016}]{wel16}
{Weltevrede} P.,  2016, \mn@doi [A\&A] {10.1051/0004-6361/201527950}, \href
  {http://adsabs.harvard.edu/abs/2016A%26A...590A.109W} {590, A109}

\bibitem[\protect\citeauthoryear{{Weltevrede}, {Edwards}  \&
  {Stappers}}{{Weltevrede} et~al.}{2006}]{wes06}
{Weltevrede} P.,  {Edwards} R.~T.,   {Stappers} B.~W.,  2006, \mn@doi [A\&A]
  {10.1051/0004-6361:20053088}, \href
  {https://ui.adsabs.harvard.edu/abs/2006A&A...445..243W} {445, 243}

\bibitem[\protect\citeauthoryear{{Weltevrede}, {Stappers}  \&
  {Edwards}}{{Weltevrede} et~al.}{2007}]{wse07}
{Weltevrede} P.,  {Stappers} B.~W.,   {Edwards} R.~T.,  2007, \mn@doi [A\&A]
  {10.1051/0004-6361:20066855}, \href
  {https://ui.adsabs.harvard.edu/abs/2007A&A...469..607W} {469, 607}

\bibitem[\protect\citeauthoryear{{Wen}, {Wang}, {Yuan}, {Yan}, {Manchester},
  {Yuen}  \& {Gajjar}}{{Wen} et~al.}{2016}]{wwy+16}
{Wen} Z.~G.,  {Wang} N.,  {Yuan} J.~P.,  {Yan} W.~M.,  {Manchester} R.~N.,
  {Yuen} R.,   {Gajjar} V.,  2016, \mn@doi [A\&A]
  {10.1051/0004-6361/201628214}, \href
  {https://ui.adsabs.harvard.edu/abs/2016A&A...592A.127W} {592, A127}

\bibitem[\protect\citeauthoryear{Wright \& Fowler}{Wright \&
  Fowler}{1981}]{wf81}
Wright G.~A.,  Fowler L.~A.,  1981, A\&A, 101, 356

\bibitem[\protect\citeauthoryear{{Yan} et~al.,}{{Yan} et~al.}{2011}]{ymv+11}
{Yan} W.~M.,  et~al., 2011, \mn@doi [MNRAS] {10.1111/j.1365-2966.2011.18522.x},
  \href {http://adsabs.harvard.edu/abs/2011MNRAS.414.2087Y} {414, 2087}

\bibitem[\protect\citeauthoryear{{Yan}, {Manchester}, {Wang}, {Yuan}, {Wen}  \&
  {Lee}}{{Yan} et~al.}{2019}]{ymw+19}
{Yan} W.~M.,  {Manchester} R.~N.,  {Wang} N.,  {Yuan} J.~P.,  {Wen} Z.~G.,
  {Lee} K.~J.,  2019, \mn@doi [MNRAS] {10.1093/mnras/stz650}, \href
  {https://ui.adsabs.harvard.edu/abs/2019MNRAS.485.3241Y} {485, 3241}

\bibitem[\protect\citeauthoryear{{Yan}, {Manchester}, {Wang}, {Wen}, {Yuan},
  {Lee}  \& {Chen}}{{Yan} et~al.}{2020}]{ymw+20}
{Yan} W.~M.,  {Manchester} R.~N.,  {Wang} N.,  {Wen} Z.~G.,  {Yuan} J.~P.,
  {Lee} K.~J.,   {Chen} J.~L.,  2020, \mn@doi [MNRAS] {10.1093/mnras/stz3399},
  \href {https://ui.adsabs.harvard.edu/abs/2020MNRAS.491.4634Y} {491, 4634}

\bibitem[\protect\citeauthoryear{{Yuen}}{{Yuen}}{2019}]{yue19}
{Yuen} R.,  2019, \mn@doi [MNRAS] {10.1093/mnras/stz951}, \href
  {https://ui.adsabs.harvard.edu/abs/2019MNRAS.486.2011Y} {486, 2011}

\makeatother
\end{thebibliography}

%\begin{thebibliography}{99}
%\end{thebibliography}

%%%%%%%%%%%%%%%%%%%%%%%%%%%%%%%%%%%%%%%%%%%%%%%%%%

%%%%%%%%%%%%%%%%% APPENDICES %%%%%%%%%%%%%%%%%%%%%

%\appendix


%%%%%%%%%%%%%%%%%%%%%%%%%%%%%%%%%%%%%%%%%%%%%%%%%% 

% Don't change these lines
\bsp	% typesetting comment
\label{lastpage}
\end{document}